\documentclass[superscriptaddress,showpacs,twocolumn,notitlepage,nofootinbib,aps,pre, 10pt]{revtex4-2}
\usepackage[normalem]{ulem}
\usepackage{graphicx}
\usepackage{color}
\usepackage{xcolor}
\usepackage{verbatim}
\usepackage{subfigure}
\usepackage{amssymb}
\usepackage{amsmath}
\usepackage{amsfonts}
\usepackage{comment}
\usepackage{chngcntr}
\usepackage{appendix}
\usepackage{lipsum}
\usepackage{dsfont}
\usepackage{xfrac}
\usepackage[makeroom]{cancel}
\usepackage{makecell}
\usepackage{lipsum}
\usepackage{tikz}
\usepackage{circledsteps}
\usepackage{CJKutf8}
\usepackage{soul}

\newcommand{\beeta}{\boldsymbol{\eta}}

\newcommand{\dd}{\text{d}}

\newcommand{\ee}{\text{e}}

\newcommand{\p}{\partial}

\newcommand{\bphi}{\boldsymbol{\phi}}

\DeclareMathAlphabet\mathbfcal{OMS}{cmsy}{b}{n}

\newcommand{\nocontentsline}[3]{}
\let\origcontentsline\addcontentsline
\newcommand\stoptoc{\let\addcontentsline\nocontentsline}
\newcommand\resumetoc{\let\addcontentsline\origcontentsline}

\usepackage{booktabs}
\usepackage{pifont}
\usepackage{xcolor}

\usepackage{amsthm}
\usepackage{bm}
\usepackage[hang, flush margin]{footmisc}
\usepackage[colorlinks=true, linkcolor=blue, citecolor=blue]{hyperref}
\usepackage{cleveref}
\crefname{equation}{Eq.}{Eqs.} 
\crefrangelabelformat{equation}{(#3#1#4--#5#2#6)}

\ExplSyntaxOn
\NewDocumentCommand{\eqrefs}{m}
 {
  \joansola_eqrefs:n {#1 }
 }

\seq_new:N \l__joansola_eqrefs_in_seq
\seq_new:N \l__joansola_eqrefs_out_seq

\cs_new_protected:Nn \joansola_eqrefs:n
 {
  \seq_set_split:Nnn \l__joansola_eqrefs_in_seq { , } { #1 }
  \seq_set_map:NNn \l__joansola_eqrefs_out_seq \l__joansola_eqrefs_in_seq
   {
    \exp_not:n {~\ref{##1} }
   }
  \int_compare:nTF { \seq_count:N \l__joansola_eqrefs_out_seq < 2 }
   { Eq.\nobreakspace }
   { Eqs.\nobreakspace }
  \textup{(\seq_use:Nn \l__joansola_eqrefs_out_seq {,~})}
 }
\ExplSyntaxOff

\newcommand{\tocless}[2]{\bgroup\let\addcontentsline=\nocontentsline#1{#2}\egroup}

\begin{document}

\title{Short-term plasticity recalls forgotten memories through a trampoline mechanism}
\author{Martina Del Gaudio}
\affiliation{Institute for Computational and Mathematical Engineering, Stanford University, Stanford, CA 94305, USA}
\author{Federico Ghimenti}
\affiliation{Department of Applied Physics, Stanford University, Stanford, CA 94305, USA}
\author{Surya Ganguli}
\affiliation{Department of Applied Physics, Stanford University, Stanford, CA 94305, USA}

\begin{abstract}
We analyze continuous Hopfield associative memories augmented by additional, rapid short-term associative synaptic plasticity. Through the cavity method, we determine the boundary between the retrieval and forgetting, or spin-glass, phase of the network as a function of the fraction of stored memories and the neuronal gain. We find that short-term synaptic plasticity yields marginal improvements in critical memory capacity. However, through dynamical mean field theory, backed by extensive numerical simulations, we find that short-term synaptic plasticity has a dramatic impact on memory retrieval above the critical capacity. When short-term synaptic plasticity is turned on, the combined neuronal and synaptic dynamics descend a high-dimensional energy landscape over both neurons and synapses. The energy landscape over neurons alone is thus dynamic, and is lowered in the vicinity of neuronal patterns recently visited by the network, just like the surface of a trampoline is lowered in the vicinity of regions recently visited by a heavy ball. This trampoline-like reactivity of the neuronal energy landscape to short-term plasticity in synapses can lead to the recall of stored memories that would otherwise have been forgotten. This occurs because the dynamics without short-term plasticity transiently moves toward a stored memory before departing from it. Thus, short-term plasticity, operating during the transient, lowers the energy in the vicinity of the stored memory, eventually trapping the combined neuronal and synaptic dynamics at a fixed point close to the stored memory. In this manner, short-term plasticity enables the recall of memories that would otherwise be forgotten, by trapping transients that would otherwise escape. We furthermore find an optimal time constant for short-term synaptic plasticity, matched to the transient dynamics, that empowers the recall of forgotten memories.
\end{abstract}
\maketitle
\section{Introduction}

At its core, a neural network is made up of a set of computational units, the neurons, and the connection mediating the influence of a neuron's activity onto another one, the synapses. Given these ingredients, two paradigmatic lines of research have been explored across the years. A first line of inquiry focuses on the types of computation and the nature of the dynamics of the neurons for a given synaptic structure~\cite{gao2015simplicity,vyas2020computation}, which is held fixed throughout the deployment of the neurons' dynamics. The seminal works on associative memories~\cite{little1974existence, hopfield1982, hopfield1984, amit1985, amit1987statistical, gardner1988optimal} and their dense versions~\cite{abbott1987storage, gardner1987multiconnected, horn1988capacities} belong to this line of inquiry. Another approach consists instead in considering synapses and neurons as a composite dynamical system, where both elements coevolve on similar timescales, with the dynamics of the synapses influencing the dynamics of the neurons, and vice versa. In biological networks, the existence of short-term synaptic plasticity driven by pre-synaptic activity is well documented~\cite{zucker2002short,abbott2004synaptic}, but an expanding body of work demonstrates that a form of short-time \textit{associative} synaptic plasticity, depending on both pre- and post- synaptic neural activity, is also possible~\cite{malenka1991postsynaptic, volianskis2013different}. This form of short-term synaptic plasticity is believed to play a crucial role in the development of working memory~\cite{lansner2023fast}. For artificial neural networks, recent work has shown that allowing fluctuations of the weights of a trained architecture can improve their performance~\cite{hinton1987using, miconi2018differentiable, miconi2020backpropamine, miconi2023learning}, and it has been reported that a combination of astrocyte-mediated higher order interactions and short-term synaptic plasticity improves the retrieval capabilities of biologically plausible associative memories~\cite{kozachkov2025neuron}. Also, interestingly, such short-term plasticity has recently been demonstrated in quantum optical realizations of Hopfield associative memories~\cite{marsh2021enhancing, marsh2025high}. Understanding limits and possibilities enabled by coupled neural-synaptic computation is thus a much desirable goal for improving both our modeling capabilities of biological networks and the performance of artificial ones. 
 
To our knowledge, Dong and Hopfield~\cite{dong1992dynamic} were the first to consider a model recurrent neural network endowed exclusively with short-term synaptic plasticity in 1992, and showed how the latter can give rise to simple forms of connectivity structure under an external stimulus. Recent work by Clark and Abbott~\cite{clark2024dynamic} showed that, when short-term associative synaptic plasticity is added on top of a random connectivity matrix, a variety of interesting dynamical regimes can be obtained, including the possibility to freeze the otherwise chaotic dynamics of the network into a given configuration. Moreover, it has been shown that superimposing short-term associative synaptic plasticity on a random connectivity background allows, after an external time-dependent drive is applied, to retain simple dynamical patterns for relatively large times~\cite{wakhloo2025associative}. In these works, the impact of short-term synaptic plasticity on a flat, or highly unstructured, connectivity background is considered. However, in both artificial and biological neural networks, short-term synaptic plasticity may act upon a structured connectivity matrix, shaped in the past by a much slower dynamical process, such as training algorithms for artificial neural network or long-term plastic development in the brain~\cite{morris2003long,lynch2004long}. We currently lack a framework to understand the interplay between long-term and short-term learning in neural computation. 

\begin{figure*}
    \includegraphics[width=\linewidth]{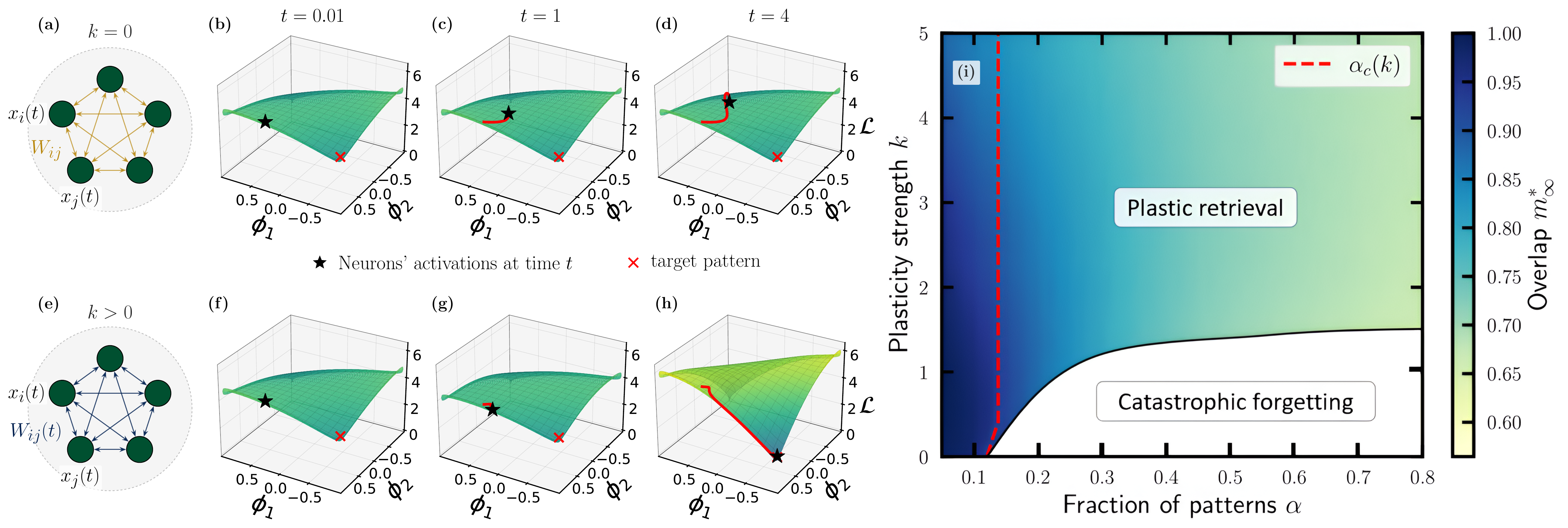}  
    \caption{Panel (a): schematic representation of a standard associative memory. The potentials of the neurons $x_i(t)$ are dynamical variables, while the synaptic weights $W_{ij}$, which contains the patterns stored by the network and mediate the interactions among different neurons, are constant. Panels (b-d): Illustration of transient retrieval and eventual catastrophic forgetting in a standard associative memory. The three snapshots show the temporal evolution of the activity pattern (black star, given by $\{\phi_1,\,\phi_2\}$, with $\phi_i\equiv \phi(x_i)\equiv \tanh \gamma x_i$) of a network made by $2$ neurons within the energy landscape $\mathcal{L} \equiv \mathcal{L}(\phi_1,\phi_2;W_{12})$, when initialized in the vicinity of a target pattern (red cross). The neural activity initially gets closer to the target pattern, but it eventually escapes away from the target's neighborhood. Panel (e): schematic representation of an associative memory with short term synaptic plasticity studied in this work. In this case, both the neurons' potentials and the synapses are dynamical variables. Panels (f-h): short-term synaptic plasticity unlocks a trampoline mechanism, which empowers pattern retrieval. The three panels shows three snapshots of the activity pattern of a network of two neurons with short-term synaptic plasticity. As discussed in Sec.~\ref{sec:global_minima}, the energy surface $\mathcal{L} \equiv \mathcal{L}(\phi_1,\phi_2;W_{12}(t))$, over the possible network's activity patterns for a given instantaneous value of the evolving synaptic weights $W_{ij}(t)$, is a dynamical quantity, which depends on the current value of the time-dependent synaptic couplings $W_{ij}(t)$. This dynamical energy landscape is lowered in regions recently visited by the neuron's activity pattern. The latter is then trapped in the vicinity of the target memory, thus enabling retrieval in a situation where the standard network is led to forgetting. Panel (i): phase diagram for the retrieval capabilities of a continuous Hopfield associative memory with short-term associative synaptic plasticity, obtained from our analysis in Sec.~\ref{sec:global_minima} and Sec.~\ref{sec:dynamics}. The heatmap shows the long-time overlap of the neuron's output with a target pattern, $m^*_\infty$, when the network is initialized in its vicinity. The long-time overlap is measured using dynamical mean-field theory, for different values of the fraction of stored pattern $\alpha$, and the short-term plasticity strength $k$. The red dashed line denotes the phase boundary, $\alpha_c(k)$, between the retrieval phase of the network and a forgetting phase, as obtained from a static analysis of the network behavior. When $\alpha>\alpha_c(k)$, without synaptic plasticity or for moderate values of $k$, the long time overlap of the network with the target pattern decays to zero. This region is where \textit{catastrophic forgetting} takes place (white region). However, as the short-term plasticity strength is increased, we report the occurrence of \textit{plastic retrieval}: long-time recovery of the target pattern takes place even when the network is loaded above its critical capacity, $\alpha>\alpha_c(k)$. The neural gain is set to be $\gamma=3.4$. Details on how panels (b-d), (f-h) were realized are given in App.~\ref{app:fig1_details}.}      \label{fig:fig1_retrieval}
\end{figure*}
How does short-term synaptic plasticity operate when superimposed on a fixed, structured connectivity background? This is the question addressed by the present work. As a minimal model neural network with structured connectivity background, we consider a recurrent neural network (RNN) with pairwise, fixed Hopfield couplings, to which we superimpose a dynamically evolving, plastic synaptic matrix. Hopfield networks constitute a celebrated model for the implementation of associative memories~\cite{little1974existence, hopfield1982, hopfield1984,ramsauer2020hopfield, krotov2023new}. Drawing inspiration from the Hebbian learning rule~\cite{hebb2005organization}, these models store in their connectivity matrix a certain number of patterns. Already in the absence of short-term synaptic plasticity, the dynamics of the neurons in the network is remarkably complicated~\cite{amit1985, amit1987statistical, Kuhn1993}, and it depends on the fraction of pattern stored in the connectivity matrix with respect to the number of neurons. If this fraction is below a given threshold, termed the critical capacity, different patterns partition the phase space of the network into distinct attractors. In this \textit{retrieval phase}, robust recovery of a given pattern is thus possible at long times, provided that the network shares some degree of alignment (measured by the so-called overlap) with the target pattern in its initial state. As the fraction of patterns exceeds the critical capacity, different basins of attraction interfere with each other, giving rise to spurious states. Even when initialized in the vicinity of a target pattern, the network is driven toward these spurious states, which bear no overlap with the target pattern. In this  \textit{catastrophic forgetting} phase, spurious basins organize into a hierarchical structure, typical of spin-glass models. The dynamics of the network in this phase is rather subtle: even though at long times the overlap with the target pattern decays to zero, the dynamical overlap exhibits a nonmonotonic behavior at intermediate times,~\cite{clark2025transient}. This \textit{transient retrieval} is a reflection of the complex structure of the basins of attraction of the network (see Fig.~\ref{fig:fig1_retrieval}(a-d) for a schematic representation of the network and a visual illustration of transient retrieval in a small network composed by two neurons). 

In this work, we study the impact of short term, associative synaptic plasticity on a Hopfield-RNN. Our salient results are summarized in Fig.~\ref{fig:fig1_retrieval}. We show that the combined neural-synaptic dynamics of the network descends over a high-dimensional, composite energy landscape, over both neurons and synapses. The energy landscape of the neurons alone is constantly reshaped by short-term synaptic activity, very much like a trampoline is constantly deformed by a ball rolling through it. The short-term plasticity feedback can thus lower the energy in the vicinity of the target pattern, eventually trapping the network to a fixed point which shares a significant overlap with the desired memory (such a trampoline mechanism is schematically illustrated in Fig.~\ref{fig:fig1_retrieval} (f-h) for a small network of two neurons). By employing a combination of static cavity method, dynamical mean field theory and extensive numerical simulations, we construct a phase diagram, displayed in Fig. \ref{fig:fig1_retrieval} (i), which describes how short-term synaptic plasticity affects the retrieval capabilities of the network. We find that, when the network has a single fixed point in the vicinity of each pattern (the so-called retrieval phase), the presence of short-term associative synaptic plasticity effectively acts as a self-coupling, whose strength is determined by the activity of the network itself, because of the plastic feedback. From a static point of view, this self-coupling leads to a marginal increase in the estimated critical capacity of the neural network, as shown by the red dashed line in Fig.~\ref{fig:fig1_retrieval} (i). However, short-term synaptic plasticity has far richer consequences when the full dynamics of the network is taken into account. Indeed, we discover that synaptic plasticity can exploit transient memory retrieval at the early times of the dynamics to  converge to a fixed point in the vicinity of a target pattern, granting robust recovery. This finding indicates that synaptic plasticity can significantly increase the size of the basins of attractions of metastable states around the memories of the network, an effect which goes unnoticed under the conventional static analysis. We christen such a dynamical, synapses-enabled recovery \textit{plastic retrieval}. Furthermore, for a given level of synaptic plasticity strength, we find that there is an optimal timescale that maximizes the impact of plastic retrieval,  suggesting the presence of a complicated interplay between the shape of the basins of attraction and the timescales of dynamical evolution of the system. Our work sheds light on the interplay between short-term and long-term memory in neural computation, and opens the way to principled applications of enabling weight fluctuations in associative memories and biologically-plausible learning. 

Our work is organized as follows: in Section~\ref{sec:model}, we introduce our model of recurrent neural network with long-term Hebbian learning rule and short-term associative synaptic plasticity, in Sec.~\ref{sec:global_minima}, we employ a static cavity method to determine the phase boundary between the retrieval and forgetting phase as a function of the fraction of stored patterns and the gain of the post-synaptic nonlinearity. In Section~\ref{sec:dynamics}, we study the dynamics of the network in the forgetting phase, using a combination of dynamical mean field theory and extensive numerical simulations. We also discuss the nature of the linear excitations around the fixed point reached by the network when loaded above its critical capacity, and we discuss the optimal time scale for which short-term synaptic plasticity maximizes the degree of pattern retrieval. We conclude this work by providing perspectives and future directions in Section~\ref{sec:conclusion}.

\section{Model and dynamics}\label{sec:model}
We consider a recurrent neural network of $N$ neurons in continuous time, following the graded-response model~\cite{Kuhn1991statistical, Kuhn1993}. Each neuron $i$ is described by the time evolution of its membrane potential, $x_i(t)$. The latter evolves according to the following leaky-integrator equation of motion,
\begin{equation}
\label{eq:neuron_dyn}
\partial_{t}x_{i}(t)=-x_{i}(t)+\sum_{j=1}^{N}W_{ij}(t)\phi(x_{j}(t))\,,
\end{equation}
where time is measured in units of the characteristic relaxation time-scale of each neuron. Each neuron synchronously produces an output $\phi(x_i(t)) \equiv \phi_i(t)$, which affects the dynamics of other neurons in the network by means of the connectivity matrix $W_{ij}(t)$. The neural output $\phi_i$  models the firing rate of the neuron, and it is given by
\begin{equation}\label{eq:phi}
\phi(x_i) \equiv \tanh(\gamma x_i)
\end{equation}
The parameter $\gamma$ is the neural gain, which controls the steepness of the non-linearity. High and low values of $\gamma$ approximate a binary neuron and a linear activation function, respectively.

The total synaptic weight matrix $W_{ij}(t)$ is at the core of our model, and it is composed by two distinct contribution:
\begin{equation}\label{eq:W}
W_{ij}(t) = J_{ij} + A_{ij}(t)
\end{equation}
The first term, $J_{ij}$, represents a static connectivity matrix, modeling a form of long-term memory process. In this work, we consider a long-term connectivity matrix $J_{ij}$ that implements an Hebbian learning rule~\cite{hebb1949, hopfield1982, amit1985}: $J_{ij}$ is a Wishart matrix storing $P = \alpha N$  $N$-dimensional memory patterns $\{\xi_i^\mu\}_{\mu=1}^P$, namely
\begin{equation}\label{eq:J_ij}
J_{ij} = \frac{1}{N}\sum_{\mu=1}^{P}\xi_{i}^{\mu}\xi_{j}^{\mu}\,.
\end{equation}
This matrix is fixed in time and represents the underlying learned structure of the network. The entries $\xi_i^\mu \in \{\pm 1\}$ are independent, identically distributed random variables with uniform distribution. The fraction of stored patterns $\alpha$ is of order $O(1)$, and represents the load of the network.

The second term on the right hand side of Eq.~\eqref{eq:W}, $A_{ij}(t)$, represents a dynamic, short-term plasticity matrix, inspired by models of activity-dependent synapses~\cite{dong1992dynamic, tsodyks1997, mongillo2008, clark2024dynamic}. It evolves on a plasticity timescale $p$, according to its own Hebbian rule, tracking the \emph{current} neural activity rather than a set of fixed stored patterns. The dynamics of $A_{ij}(t)$ reads
\begin{equation}
\label{eq:stp_dyn}
p\partial_{t}A_{ij}(t)=-A_{ij}(t) + \frac{k}{N}\phi_{i}(t)\phi_{j}(t)\,.
\end{equation}
The two contributions on the right hand side represent respectively a linear decay term that eventually leads past activity to be washed out from $A_{ij}(t)$, and a synaptic reinforcement term that strengthens the connections among neurons that are simultaneously firing. The plasticity strength $k$ sets the intensity of this short-term reinforcement effect. This model thus couples the neural dynamics in Eq.~\eqref{eq:neuron_dyn} with the synaptic dynamics in Eq.~\eqref{eq:stp_dyn}, allowing us to study the interplay between fixed-pattern storage and activity-dependent synaptic modulation. If we were to choose the long-term connectivity matrix $J_{ij}$ to be a random matrix with independent Gaussian entries, we would recover the model studied by Clark and Abbot~\cite{clark2024dynamic}. For $J_{ij} =0$, we recover instead the model with purely short-term synaptic plasticity, studied originally by Dong and Hopfield~\cite{dong1992dynamic}.

In the absence of short-term synaptic plasticity, for $k=0$, the salient features of the model in Eq.~\eqref{eq:neuron_dyn} are well known~\cite{Kuhn1991statistical, Kuhn1993}. The most relevant parameter describing the retrieval capabilities of the network is the overlap $m^{\mu^*}(t)$ with respect to a target, or \textit{condensed}, pattern $\xi^{\mu^*}_i$, defined as
\begin{equation}
    m^{\mu^*}(t) \equiv \frac{1}{N} \sum_{i=1}^N \left\langle \phi_i(t)\xi_i^{\mu^*}\right\rangle,\, 
\end{equation}
where the average $\langle\cdot\rangle$ denotes an average over all the realizations of the patterns $\{\xi_i^\mu\}$ stored in $J_{ij}$. The long time value of $m^{\mu^*}(t)$ describes the degree of alignment of the network with the target pattern: a zero value of the overlap indicates that the pattern has been forgotten, while a value of $m^{\mu^*}(t)$ close to unity  indicates successful retrieval. For a given value of the neural gain $\gamma$, there is a critical fraction of patterns $\alpha_c(\gamma)$ that can be robustly retrieved by the network. For $\alpha<\alpha_c(\gamma)$, if the network is allowed to evolve starting from  the vicinity of a target pattern $\xi^{\mu^*}$, then the long time value of its overlap with the target pattern is a finite quantity, of order $O(1)$. The phase space of the model is partitioned into different basins, associated to the different patterns. This is the \textit{retrieval} phase. For $\alpha>\alpha_c(\gamma)$, the long time value of the overlap with the target pattern approaches zero as the size of the system goes to infinity. This is the \textit{catastrophic forgetting}  phase, and the phase space of the system has a spin-glass structure, with an exponentially large number of energy minima, that are \textit{unaligned} with respect to any of the stored patterns. The transition from the retrieval to the forgetting phase is driven by the noise produced by the uncondensed patterns, which give rise to spurious states that do not overlap with the stored patterns and that can trap the system. When the neuronal dynamic starts near a desired memory, it eventually escapes the vicinity of the desired memory at late times. This occurs because the energy landscape governing the descent dynamics no longer has local minima with large basins near the desired memories. Our investigation thus starts by looking at the properties of the network at the fixed points of its dynamics, and by asking  how is the transition from the retrieval to the forgetting phase affected by the presence of short-term synaptic plasticity.

\section{Retrieval phase with short-term synaptic plasticity}\label{sec:global_minima}

We start by observing that the coupled neurons-synapses dynamics, \cref{eq:neuron_dyn,eq:phi,eq:W,eq:J_ij,eq:stp_dyn}, minimizes a Lyapunov (or energy) function $\mathcal{L}\left[\{\phi_i\}_{i=1}^N,\,\{A_{ij}\}_{i,j=1}^N\right]$ which is defined, up to an additive constant, as
\begin{align}\label{eq:Lyapunov}
    \begin{split}
        \mathcal{L}&\left[\{\phi_i\}_{i=1}^N,\,\{A_{ij}\}_{i,j=1}^N\right]
        \equiv -\frac{1}{2}\sum_{i,j=1}^N W_{ij}\,\phi_i\phi_j
        \\&+ \sum_{i=1}^N \int_0^{\phi_i} \phi^{-1}(v)\,\dd v
        + \frac{N}{4k}\sum_{i,j=1}^NA_{ij}^2\,.
    \end{split}
\end{align}
Indeed, we show in App.~\ref{app:Lyapunov} that 
\begin{equation}\label{eq:dLyapunov_dt}
    \frac{\dd\mathcal{L}}{\dd t} \leq 0\,,
\end{equation}
and that the inequality is saturated only at the fixed points of the dynamics of the network. The first term in Eq.~\eqref{eq:Lyapunov} involves the coupling energy among the different neurons' outputs through the connectivity matrix $W_{ij}$. The second term incorporates the membrane potential leak of each neuron, and the last term is a quadratic potential that confines the short-term plastic matrix $A_{ij}$. If $J_{ij}=0$ (see Eq.~\eqref{eq:W}), then we recover the purely plastic case considered by Dong and Hopfield~\cite{dong1992dynamic}, while in the limit $k\to 0$, we recover the energy of a recurrent neural network with purely long-term Hebbian plasticity studied by K\"uhn~\cite{Kuhn1991statistical, Kuhn1993}. A neural network with short-term synaptic plasticity and Hebbian learning thus evolves in a high-dimensional energy landscape over an extended set of degrees of freedom. If we consider the restriction of the energy landscape $\mathcal{L}[\{\phi_i\}_{i=1}^N;\,\{W_{ij}(t)\}_{i,j=1}^N]$ to the neurons' activity alone, for a given instantaneous value of the synaptic weights $W_{ij}(t)$, we see that such a landscape is constantly reshaped by the evolution of short-term synaptic plasticity, as the dynamics unfolds. The energy of recently visited neuronal configurations is lowered by means of the evolution of the short-term plasticity matrix $A_{ij}$. As mentioned in the Introduction, this mechanism is akin to a trampoline being deformed by a ball rolling through it. As it will be seen in Sec.~\ref{sec:dmft}, such a feedback mechanism between neural and synaptic activity is crucial in affecting the retrieval capabilities of the network.      

The properties of the network at the fixed point of the dynamics, \cref{eq:neuron_dyn,eq:phi,eq:W,eq:J_ij,eq:stp_dyn}, can be studied using the static cavity method~\cite{mezard1987,del2014cavity}, originally introduced in the study of spin-glass systems. The derivation is detailed in App.~\ref{app:cavity_static}, and it allows us to reduce the system of $N + N^2$ equations for the fixed points of the dynamics of the full neuronal network to a one-dimensional problem for the effective fixed point condition of a tagged neuron in the system. The effect of short-term synaptic plasticity, as well as the influence from the other neurons, can be integrated out from the fixed point equations, and their influence can be described by means of a set of order parameters that, as customary in a mean-field setting, are self-consistently determined from the properties of the population of neurons at the fixed point itself. In particular, the cavity method allows to determine the overlap $m^*$ of the network with a target, condensed pattern at the fixed point, as
\begin{equation}\label{eq:cavity_overlap}
    m^* = \left\langle \xi^* \phi \right\rangle_{z,\xi^*}\,,
\end{equation}
where the output $\phi$ is the output of the tagged neuron in the network at the fixed point, and satisfies the equation 
\begin{equation}\label{eq:cavity_phi_main}
\phi = g\left(\xi^* m^* + \sigma z + \Gamma\phi\right)\,,
\end{equation}
where $g(x) \equiv \tanh \gamma x$, and $\xi^*$ is a discrete random variable, sampled  uniformly from the set $\{-1,1\}$, representing a mean-field description of one of the condensed  patterns. The scalar $z$ is a Gaussian noise of mean zero and unit variance, which represents the cumulative influence of all the other neurons on the cavity neuron. The average $\langle\cdot\rangle_{z,\xi^*}$ is an average over the realizations of the Gaussian noise and the pattern $\xi^*$. Averages over different realizations of the noise $z$ and the target pattern $\xi^*$ are statistically equivalent to averages over the population of neurons at the fixed point, under different statistical realizations of the $N$-dimensional patterns $\{\xi_i^\mu\}$. The strength of the fluctuations of the noise $z$ is set by the variance $\sigma^2$. The third term contributing to the argument of $g$ in Eq.~\eqref{eq:cavity_phi_main} describes instead the backpropagated response of the system to the cavity neuron. It is proportional to the neuron output, $\phi$ and it is screened by a constant $\Gamma$. The variance $\sigma^2$ and the screening factor $\Gamma$ read, respectively,
\begin{align}\label{eq:sigma2_Gamma}
    \begin{split}
         \sigma^2 &= \frac{\alpha q}{\left[1 - \overline{\chi}^{(\phi,h)}\right]^2}\\
         \Gamma &= \frac{\alpha}{1 - \overline\chi^{(\phi,h)}} + kq\,,
    \end{split}
\end{align}
The quantity $q$ is the mean squared activity of the population of neurons at the fixed point, and it is determined by the self-consistent relation
\begin{equation}\label{eq:q}
    q = \left\langle \phi^2 \right\rangle_{z,\xi^*}\,,
\end{equation}
while $\overline{\chi}^{(\phi,h)}$ is the static susceptibility of the output of a neuron in the network at the fixed point, under a small perturbation to its membrane potential. It satisfies the self-consistent equation
\begin{equation}\label{eq:cavity_chi}
    \overline\chi^{(\phi,h)}
     = \frac{\left\langle \frac{\dd \phi}{\dd z}  \right\rangle_{z,\xi^*}}{\sqrt{\alpha q} + \left\langle \frac{\dd \phi}{\dd z}  \right\rangle_{z,\xi^*}}\,.
\end{equation}
The expressions of the screening constant $\Gamma$, of the noise variance $\sigma^2$, and of the static susceptibility $\overline{\chi}^{(\phi,h)}$ stem from a feedback mechanism between the network's output and its overlap with all the non-targeted patterns along which the network does not align. Changes in the neuron's output produce shifts in the uncondensed overlaps, which in turn backpropagate into the dynamics of the neurons. At the fixed point, this feedback mechanism produces a screening effect, which once taken into account, yields the self-consistent relations described by \cref{eq:sigma2_Gamma,eq:cavity_chi}.

The properties of the fixed point determined above depend on the realizations on the noise $z$. The statistics of this noise is taken to be Gaussian, but this assumption can break down due to the development of a nonlinear instability of the network against small random perturbations to the neurons' membrane potential. This nonlinear instability signals the transition to a spin-glass phase~\cite{amit1985, amit1987statistical}, and it is studied in App.~\ref{app:cavity_breakdown}. There, we derive a condition for the validity of the cavity calculation, which reads
\begin{equation}\label{eq:cavity_braking}
    \left\langle \left[\frac{\dd \phi}{\dd z}\right]^2\right\rangle_{z,\xi^*} \leq q\,.
\end{equation}
When Eq.~\eqref{eq:cavity_braking} is violated, then the terminal state of the network cannot be described by a unique fixed point within the basin of attraction of the condensed pattern. Multiple fixed-point solutions are possible, and their description requires more complex \textit{ansatz} for the statistics of the noise $z$~\cite{mezard1987, fischer1993spin}.

\cref{eq:cavity_overlap,eq:cavity_phi_main,eq:sigma2_Gamma,eq:q,eq:cavity_chi,eq:cavity_braking} describe the statistical properties of our recurrent neural network with short-term and long-term synaptic plasticity at a fixed point of the dynamics, and in particular, they allow to compute the average overlap with a target pattern.
For $k=0$, we recover the equations obtained by K\"{u}hn for a Hopfield-RNN using the replica method~\cite{Kuhn1991statistical, Kuhn1993}. On the other hand, when $k\neq 0$, we see from Eq.~\eqref{eq:sigma2_Gamma} that turning short-term synaptic plasticity on leads to a shift of the screening constant $\Gamma$. As long as fixed points of the dynamics are considered, this shift can equivalently be obtained by setting $k=0$ and increasing the diagonal elements of the matrix $J_{ij}$ in Eq.~\eqref{eq:J_ij} by a constant amount $kq$. Thus, within the validity region of the cavity method, short-term synaptic plasticity has the same effect as an additional self-coupling in a non-plastic network, with the salient aspect that the strength of the self-coupling needs to be self-consistently determined from the mean squared output $q$, given by Eq.~\eqref{eq:q}.   

\begin{figure}
    \includegraphics[width=\columnwidth]{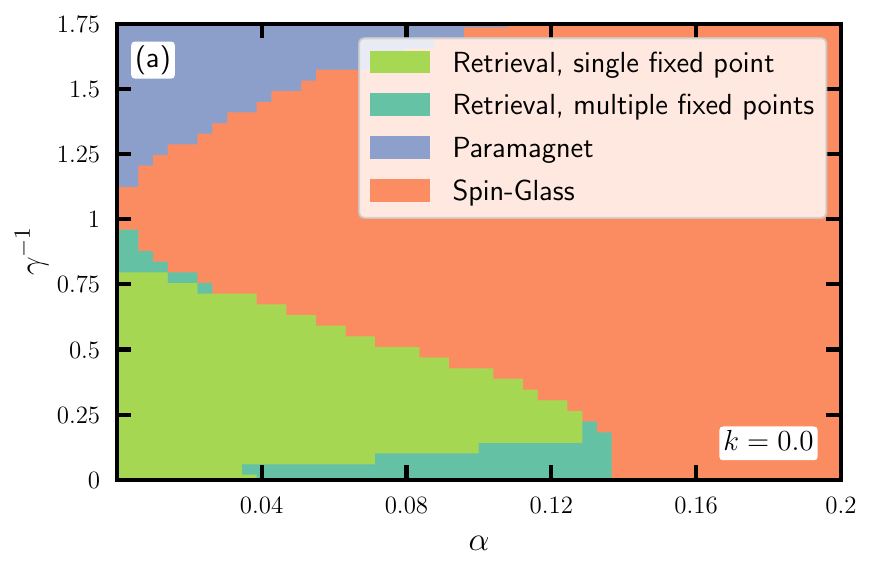} 
    \includegraphics[width=\columnwidth]{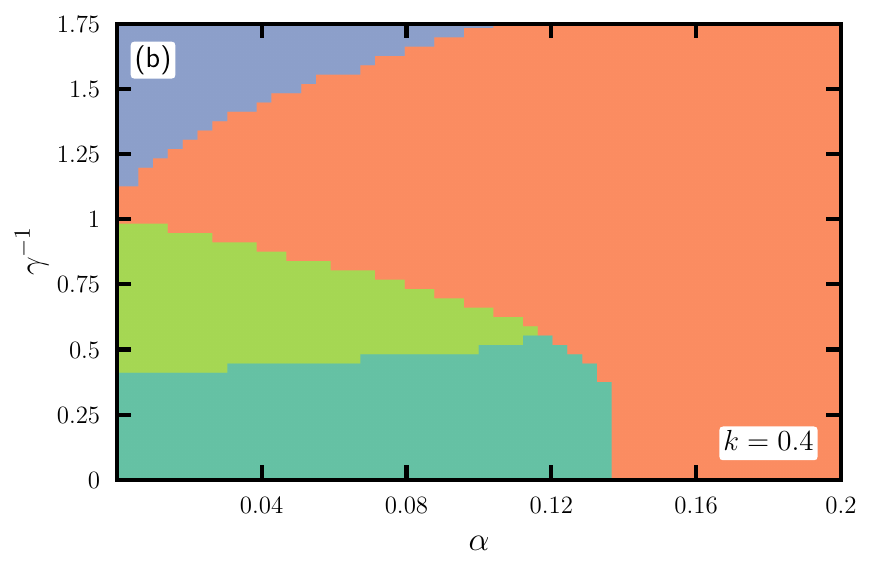}
    \caption{\textbf{Phase diagram of the fixed point of an Hopfield-RNN with short-term associative synaptic plasticity.} Different phases are obtained as the fraction of stored patterns $\alpha$ and the inverse gain $\gamma^{-1}$ are varied. We identify a retrieval phase with a single fixed point (light green region) where the overlap of the network activity with the target pattern is larger than $0$; a multiple-fixed points retrieval phase (dark green region), where the network is prone to nonlinear instabilities but it maintains a finite overlap with the target pattern. A paramagnet phase (blue region), where the network converges to a unique, quiescent, fixed point, with no overlap with the target pattern; and a spin-glass phase (orange region) with multiple fixed points and zero overlap with the target pattern. The phase diagrams are for null, and moderate values of the short-term plasticity strength ($k=0$, panel (a), and $k=0.4$, panel (b), respectively). 
    }
    \label{fig:static_phase_diagram}
\end{figure}

The fixed-point equations, \cref{eq:cavity_overlap,eq:cavity_phi_main,eq:sigma2_Gamma,eq:q,eq:cavity_chi,eq:cavity_braking}, can be solved numerically by iteration for different values of the fraction of patterns $\alpha$, output gain $\gamma$, and short-term synaptic strength $k$, by adapting the method of~\cite{Kuhn1993}. Once convergence is achieved, the value of the different order parameters can be measured, and a phase diagram for the properties of the fixed points of the dynamics of the network can be constructed, as a function of the fraction of stored patterns $\alpha$ and the inverse gain $\gamma^{-1}$, for different values of the short-term synaptic strength. These phase diagrams are illustrated in Fig.~\ref{fig:static_phase_diagram}, which are obtained respectively without short-term synaptic plasticity, Fig.~\ref{fig:static_phase_diagram} (a), and for a moderate value of the plasticity strength, Fig.~\ref{fig:static_phase_diagram} (b).
In both cases, four distinct phases can be identified. For low value of the load $\alpha$, and large enough values of the gain $\gamma$, a retrieval, single-fixed point phase (light green region in Fig.~\ref{fig:static_phase_diagram}), the overlap of the network with the condensed pattern is close to unity, $m^*\approx 1$,  and the stability condition given by Eq.~\eqref{eq:cavity_braking} is obeyed. The attractors pertaining to each pattern are distinct from each other, and each of them is endowed by a unique fixed point. They are indeed global minima of the energy function $\mathcal{L}$ in Eq.~\eqref{eq:Lyapunov}. The dark green region in Fig.~\ref{fig:static_phase_diagram} instead represents a retrieval phase with multiple fixed points. Here too, we have $m^*\approx 1$, but the stability condition in Eq.~\eqref{eq:cavity_braking} is violated. The basins of attractions around the patterns of the network are metastable, and they host multiple fixed points. Even though in this region the cavity calculation breaks down, previous works on associative memories~\cite{amit1985, amit1987, tokita1994replica} indicate that the cavity solution still yields an accurate description of the retrieval capabilities of the network. Upon increasing the fraction of patterns stored in the connectivity matrix, the system transitions toward a spin glass phase (orange region in Fig.~\ref{fig:static_phase_diagram}), where the overlap with a condensed pattern is zero, $m^*=0$, and the stability condition given by Eq.~\eqref{eq:cavity_braking} breaks down.  Finally for low value of the fraction of stored patterns and high gains, the neural network has a unique fixed point, a paramagnetic state with zero overlap with the condensed pattern, $m^*=0$, and where the network is completely quiescent (blue region in Fig.~\ref{fig:static_phase_diagram}). 

The salient result of our analysis is an understanding of the impact of short-term synaptic plasticity on the retrieval phase boundary, i.e. the curve $\alpha_c(\gamma,k)$ that separates the retrieval phases from the spin-glass phase. A comparison between Fig.~\ref{fig:static_phase_diagram} (a) and Fig.~\ref{fig:static_phase_diagram} (b) shows that short-term synaptic plasticity increases the retrieval region of the system at moderate gains $\gamma$, but yields little improvement at high gains. In particular, the maximal critical capacity that can be obtained from an Hopfield-RNN, $\alpha_c^\text{max} = \lim_{\gamma\to\infty} \alpha_c(\gamma,k=0)\approx 0.138$, is left unaffected by the presence of short-term synaptic plasticity. Indeed, the additional self-coupling term appearing in Eq.~\eqref{eq:sigma2_Gamma} becomes irrelevant in the limit $\gamma\to\infty$, as the output gain of the neuron converges to a sign function. On the other hand, the limit of infinite gain constitutes a strict upper bound to the critical capacity capacity. This follows from the boundedness of the neuron's output function in the continuous-time framework~\cite{hopfield1984, Kuhn1993}: for any finite gain, $|\tanh(\gamma x)| < 1$, whereas the infinite-gain limit yields a binary output $|\text{sgn}(x)| = 1$. Consequently, the macroscopic overlap $m^*$ for finite $\gamma$ is strictly smaller than the one obtained in the infinite-gain limit. Such a reduction in the signal strength renders the retrieval state less robust against the interference noise from uncondensed patterns. Thus, we expect to have $\alpha_c(\gamma,k) < \alpha_c^\text{max}$ for finite $\gamma$.

Inspection of Fig~\ref{fig:static_phase_diagram} (a-b) reveals that the the single-fixed point, retrieval region shrinks as short-term synaptic plasticity is turned on. Such shrinkage can be understood by inspection of the stability condition in Eq.~\eqref{eq:cavity_braking}. A sufficient condition for the breakdown of the stability of the cavity method is obtained when the fixed point equation for the neuron output, Eq.~\eqref{eq:cavity_phi_main} admits multiple solutions for some realization of the noise $z$, which results in a discontinuity in the derivative $\frac{\dd \phi}{\dd z}$. Such a situation corresponds to the opening of a gap in the distribution of the membrane potential of the population of neurons at the fixed point~\cite{Kuhn1993}. In particular, when $\sigma z=-\xi^* m^*$, Eq.~\eqref{eq:cavity_phi_main} becomes $\phi = g(\Gamma \phi)$. If the addition of a self coupling yields higher values of the screening constant $\Gamma$, then it is easier to have a set of order parameters for which multiple solutions are possible. Moreover, since $g(x) = \tanh (\gamma x)$, we see how increasing the gain $\gamma$ makes it easier for the system to develop instabilities. Short-term synaptic plasticity thus increases the reactivity of the output of the population of neurons under small changes in their membrane potential, rendering the system more prone to fluctuations.

We conclude this Section by commenting on the boundaries between the paramagnetic phase and the other phases in Fig.~\ref{fig:static_phase_diagram} (a--b). An analysis carried out in App.~\ref{app:boundary_para} allows us to determine the critical gains $\gamma_c(k)$ and $\gamma_g(\alpha,k)$, denoting the boundaries between the paramagnetic phase and the retrieval phase as $\alpha\to 0$, and the boundary between the paramagnetic phase and the spin-glass phase at finite $\alpha$, respectively. Our result reads
\begin{align}\label{eq:gamma_c_gamma_g}
    \begin{split}
        \gamma_c(k)^{-1} &= 1 \\
        \gamma_{g}(\alpha,k)^{-1} &= (1 + \sqrt{\alpha})^2\,.
    \end{split}
\end{align}
These two quantities are independent of $k$, and are, in particular, identical to their values in the absence of short-term synaptic plasticity. This situation is different from the case where a constant self-coupling term is added to the network, as studied in~\cite{Kuhn1993}. This feature is a consequence of the fact that, when short-term synaptic plasticity is turned on, the strength of the self-coupling needs to be self-consistently determined from the network activity, according to Eq.~\eqref{eq:q}. In the paramagnetic state, the network is quiescent, $\phi=0$, and therefore the self-overlap $q$ also vanishes, $q=0$, which explains the absence of dependence on $k$ in Eq.~\eqref{eq:gamma_c_gamma_g}.

The analysis carried out in this section demonstrates that short-term synaptic plasticity has a marginal effect on the width of the retrieval region of the neural network. Quite interestingly, we have shown that short-term synaptic plasticity can actually destabilize the retrieval phase. These salient features are due to the fact that short-term synaptic plasticity acts at the fixed point of the dynamics as an effective self-coupling. If the neuron nonlinearity is sufficiently steep, then the self coupling is irrelevant in changing the neurons' output. However, the analysis carried out so far faithfully tracks only the properties of the single-fixed point phases of the neural network, while ignoring the full dynamical history of the system, which becomes crucial to understand as multiple fixed-point phases appear. In the spin-glass phase, the geometry of the energy landscape $\mathcal{L}$ is certainly more complex, and the full evolution of the system has to be tracked. Thus,  to understand the impact of short-term synaptic plasticity in the forgetting phase, we turn to the study of the dynamics of the network. 

\section{Dynamical plastic retrieval}\label{sec:dynamics}
To understand the retrieval ability of the network when the static cavity method breaks down, we study the dynamics of the network when its capacity exceeds the critical threshold, employing a combination of finite-size simulations and dynamical mean-field theory (DMFT)~\cite{sompolinsky1981dynamic, sompolinsky1982relaxational, sompolinsky1988, clark2024dynamic, clark2025transient}. 
\subsection{Dynamical mean-field theory}\label{sec:dmft}

The dynamical mean field theory framework allows us to derive an effective equation for the dynamics of a tagged neuron in the network, by integrating out the influence of all the other neurons and the plastic synapses onto the tagged neuron. The Markovian, many-body dynamics in \cref{eq:neuron_dyn,eq:phi,eq:W,eq:J_ij,eq:stp_dyn} becomes an effective non-Markovian dynamics for an individual neuron. The effect of the surrounding system is encoded in a memory term and a stochastic noise. In the limit $N\to\infty$, thanks to the fully connected topology of the model, the statistics of the noise become Gaussian and colored. Moreover, both the noise autocorrelation function and the memory kernel can be self-consistently expressed in terms of the statistics of the effective single-neuron process. The dynamical evolution of network population averages, such as the overlap with a target pattern, can be obtained as averages over the effective single-neuron process under different realizations of the effective noise. 

The detailed derivation of the DMFT is presented in App.~\ref{app:dmft}. The final result is an effective single-site stochastic process describing the evolution of the membrane potential $x(t)$ of a tagged neuron in the system, which reads
\begin{align}\label{eq:dmft_main_x}
    \begin{split}
        \partial_{t}x(t) &= -x(t) + \xi^*m^*(t) + \eta(t)  \\
        & + \int_{0}^{t} \left[ \alpha R_{m}(t,\tau) + K_{\text{plast}}(t,\tau) \right] \phi(x(\tau)) \dd\tau\,.
    \end{split}
\end{align}
The first term on the right-hand side stems from the single-site decay term each neuron is subjected to. The term $\xi^* m^*(t)$ involves the dynamical overlap $m(t)$ of the network with a condensed pattern, and it plays the role of an effective signal driving the neuron toward the target pattern. The dynamics overlap is determined self-consistently as
\begin{equation}\label{eq:m*}
    m^*(t) = \langle \xi^* \phi(x(t))\rangle_{\eta,\xi^*}\,,
\end{equation}
where $\langle\cdots  \rangle_{\eta,\xi^*}$ is an average over the realizations of the mean field pattern $\xi^*$, which takes the value $\pm 1$ with uniform probability, and the noise $\eta(t)$ appearing in Eq.~\eqref{eq:dmft_main_x}. The latter stems from the (spurious) overlaps with an extensive number of uncondensed patterns. In the limit of an infinite network size, as a consequence of the central limit theorem, the statistics of the noise $\eta(t)$ is Gaussian, with zero mean and dynamical autocorrelation function given by
\begin{align}\label{eq:eta_eta}
    \begin{split}
        \langle \eta(t) \eta(t')\rangle_{\xi^*, \eta} &= \alpha \int_0^t \int_0^{t'}\dd\tau\,\dd\tau'\, R_m(t,\tau)\\
        &\times R_m(t',\tau') C_\phi(\tau,\tau')\,.
    \end{split}
\end{align}
The function $R_m(t,\tau)$ encodes the response of an overlap with an uncondensed pattern at time $t$, when subjected to a small punctuated shift at time $\tau$. This perturbation affects the outputs of the neuron in the network, which in turn propagates back to the uncondensed overlaps. As a result of this feedback loop, the response kernel $R_m(t,\tau)$ is determined by the Volterra equation
\begin{align}\label{eq:Rm}
    \begin{split}
        R_m(t,t') = \delta(t-t') + \int_{t'}^t \dd \tau R_\phi(t,\tau)R_m(\tau,t')\,,
    \end{split}
\end{align}
where the response function $R_\phi(t,t')$ is the response in the output of the tagged neuron at time $t$ when a small field $h(\tau) =\delta(\tau-t')h$ is applied at time $t'$, namely
\begin{equation}\label{eq:Rphi}
    R_\phi(t,t') = \left\langle \frac{\delta \phi(x(t))}{\delta h(t')}\right\rangle_{\eta,\xi^*}\,.
\end{equation}
A similar feedback process is responsible for the form of the autocorrelation function of the noise $\eta(t)$ in Eq.~\eqref{eq:eta_eta}, which involves the autocorrelation function of the neuron outputs $C_\phi(t,t')$, determined by the self-consistent relation
\begin{equation}\label{eq:Cphi}
    C_\phi(t,t') = \langle \phi(t)\phi(t')\rangle_{\eta,\xi^*}\,.
\end{equation}
Finally, the effect of short-term synaptic plasticity is encoded in the short-term plasticity kernel $K_\text{plast}(t,t')$, given by
\begin{equation}\label{eq:Kplast}
    K_\text{plast}(t,t') = \frac{k}{p}\ee^{-(t-t')/p}C_\phi(t,t')\Theta(t-t')\,.
\end{equation}
Such a kernel propagates the activity of a neuron in the network forward in time, from a past time $t'$ to a later time $t$, provided that the neuron is typically firing at both these times. The strength of the signal provided by the plastic kernel decays exponentially over the short-term plasticity time-scale $p$.

\cref{eq:dmft_main_x,eq:m*,eq:eta_eta,eq:Rm,eq:Rphi,eq:Cphi,eq:Kplast} describe the DMFT of a neuron in our recurrent neural network with short-term synaptic plasticity. Their validity extends across the whole $\gamma^{-1}-\alpha$ phase diagram, and can thus be used to describe the dynamical evolution of the network, even in the catastrophic forgetting phase. For $k=0$, we recover the case with purely long-term associative memory~\cite{clark2025transient}. In the presence of short-term synaptic plasticity, the plastic kernel $K_\text{plast}$ strengthens the memory term in Eq.~\eqref{eq:dmft_main_x}, propagating the activity of the neuron from the past. This is the dynamical realization of the trampoline-like effect that was commented at the beginning of Sec. \ref{sec:global_minima}, when discussing the energy landscape of the recurrent neural network. The structure of \cref{eq:dmft_main_x,eq:m*,eq:eta_eta,eq:Rm,eq:Rphi,eq:Cphi,eq:Kplast} is quite complex, and even in the absence of short-term synaptic plasticity, an analytical solution is not known. We can, however, make progress by solving \cref{eq:dmft_main_x,eq:m*,eq:eta_eta,eq:Rm,eq:Rphi,eq:Cphi,eq:Kplast} numerically. The procedure is detailed in App.~\ref{app:sim_details}. In particular, the numerical solution allows us to track the dynamical evolution of the overlap $m^*(t)$, when the network is initialized in the vicinity of a condensed pattern, $m^*(0) \neq 0$. Such a nonzero initial overlap value is achieved by tuning the distribution of the initial condition $x(0)$. Such a numerical solution is accompanied by finite-size simulations of the dynamics of a full neural-network (details in App.~\ref{app:finite_size_sim}), which we use to support the validity of the numerical integration of the DMFT equations. 

\begin{figure}                  
    \includegraphics[width=\columnwidth]{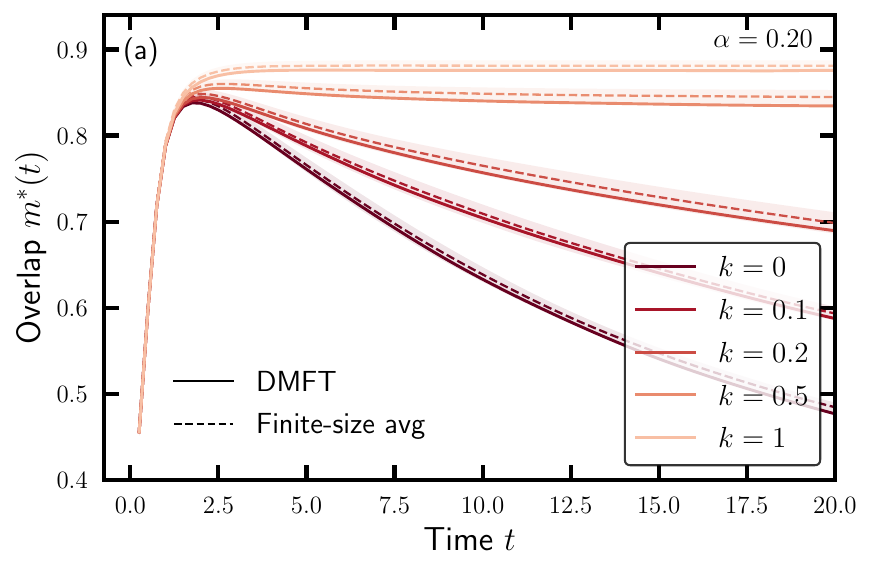} 
    \includegraphics[width=\columnwidth]{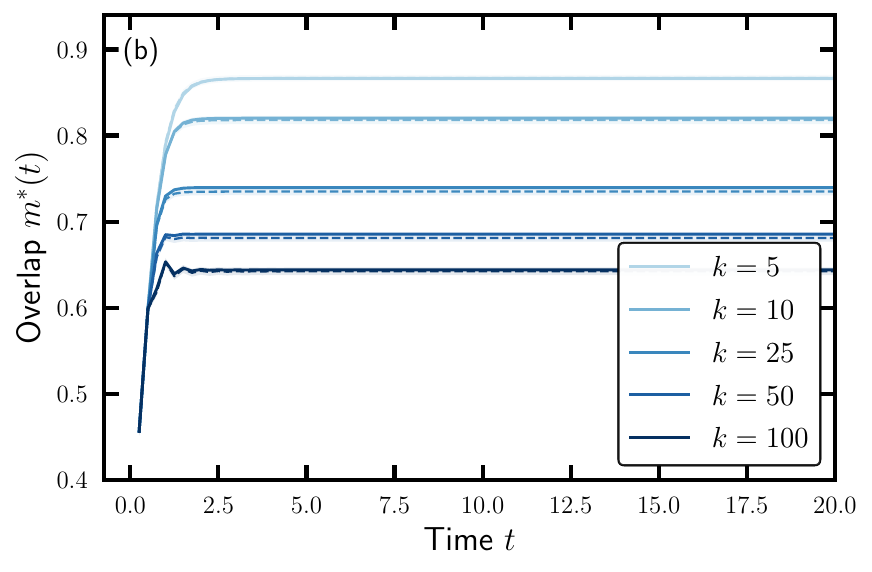} 
    \caption{\textbf{Plastic retrieval stabilizes transient memories} Temporal evolution of the overlap $m^*(t)$, as obtained from finite size simulations (dashed lines) and by numerical solution of the DMFT (solid lines). The network is initialized in the vicinity of a target pattern, with the initial overlap set to $m^*(0) \approx 0.46$, and the evolution of $m^*(t)$ for different values of the short-term plasticity strength $k$ is displayed. We have chosen the fraction of stored patterns to be above the critical capacity, $\alpha>\alpha_c(k)$. For moderate value of $k$, the retrieval of the target pattern is only a transient phenomenon. But when the plastic strength becomes larger than a critical threshold, plastic retrieval takes place, and long-time recovery of the target pattern is possible.}
    \label{fig:dmft_vs_sims}
\end{figure}

The main result obtained from the numerical solution of the DMFT equations and finite-size simulations is displayed in Fig.~\ref{fig:dmft_vs_sims}, where we plot the evolution of the condensed overlap $m^*(t)$ as a function of time, for fixed values of the capacity $\alpha$, the gain $\gamma$ and different values of the short-term synaptic plasticity strength $k$. 
The capacity is chosen above the maximal critical capacity of the system $\alpha>\alpha_c^\text{max} \approx 0.138$, so that, from the perspective of the static cavity method,  the network is in a spin-glass, forgetting phase. For $k=0$, the overlap $m^*(t)$ initially increases, reaching a maximum of height $m^*(t_\text{max})=0.82$ around a time $t_\text{max} \approx 1.5$, while at longer times, the overlap $m^*(t)$ slowly decays down to $0$. The nonmonotonic behavior of $m^*(t)$ for $k=0$ is known as \textit{transient retrieval}, and it is an indication of the complex structure of the high-dimensional basins of attraction of the Hopfield recurrent neural network above the critical capacity. The slow decay at longer times is instead a well-recognized instance of catastrophic forgetting, and it demonstrates the inability of the network to robustly store patterns above its critical capacity (see Fig~\ref{fig:long_time_finite_size} (a) in App.~\ref{app:finite_size_sim} for results from finite-size simulations at longer times). However, in the presence of short-term synaptic plasticity, the situation becomes quite different. For small values of $k$, we still observe the phenomena of transient retrieval and catastrophic forgetting, albeit the maximum value of the overlap achieved during transient retrieval increases with $k$, and the decay of the overlap occurs more slowly. However, when $k$ exceeds a critical value, which for the parameter chosen in this case is $\approx 0.8$, the overlap $m^*(t)$ plateaus to a nonzero value at large times. Interestingly, the value of the long-time plateau varies nonmonotonically with the strength of the short-term synaptic plasticity, as shown in Fig.~\ref{fig:dmft_vs_sims} (b). Moreover, the value of the long-time plateau of $m^*(t)$ for large values of $k$ can even be smaller than the maximum overlap value achieved in the absence of short-term synaptic plasticity (see Fig~\ref{fig:long_time_finite_size} in App.~\ref{app:finite_size_sim} for results from finite-size simulations at longer times, both with and without short-term synaptic plasticity). These findings demonstrate that short-term synaptic plasticity can stabilize the dynamic transient retrieval phase of a Hopfield recurrent neural network, generating new metastable states which considerably overlap with the stored patterns. These metastable states then trap the network in the vicinity of the target pattern, allowing for robust retrieval. Short-term synaptic plasticity allows past network activity to be propagated forward in time to the current neural activity, providing a memory effects that slow down the network and facilitates its arrest, hindering the escape from the target pattern and the access to the catastrophic forgetting regime. In the trampoline-like analogy put forward at the beginning of Sec.~\ref{sec:global_minima}, the continuous reshaping by short-term synaptic plasticity of the effective neural energy landscape leads to a slowdown of the network, in the same way a ball rolling through a trampoline is hindered in its motion with respect to a ball rolling over a solid surface. 

The discovery of dynamical pattern stabilization by means of short-term associative synaptic plasticity is the main result of this work. In the next Section, we give a closer look at the linear excitations around the retrieval configurations reached through, and stabilized by, short-term synaptic plasticity.       

\subsection{Stable fixed points from short-term synaptic plasticity}\label{sec:jacobian}

In this Section, we discuss the spectrum of the Jacobian of the dynamics of the Hopfield recurrent neural network with short-term synaptic plasticity, evaluated at the fixed points reached through plastic retrieval. The Jacobian $\mathcal{J}$ of the dynamics in \cref{eq:neuron_dyn,eq:phi,eq:W,eq:J_ij,eq:stp_dyn} is an $(N^2+ N)\times (N^2+N)$ matrix, which can be separated into four distinct blocks, namely
\begin{align}
\mathcal{J}
\equiv\begin{pmatrix}
\frac{\p \dot x_i}{\p x_j} & \frac{\p \dot x_{i}}{\p A_{jk}} \\
\frac{\p \dot A_{ij}}{\p x_k} & \frac{\p \dot A_{ij}}{\p A_{kl}}
\end{pmatrix}\equiv 
\begin{pmatrix}
\mathcal{J}_{xx} & \mathcal{J}_{xA} \\
\mathcal{J}_{Ax} & \mathcal{J}_{AA}
\end{pmatrix}\,,
\end{align}
The four blocks $\mathcal{J}_{xx},\,\mathcal{J}_{xA},\,\mathcal{J}_{Ax},\,\mathcal{J}_{AA}$ correspond to neuron-neuron, neuron-synapses, synapses-neuron, and synapses-synapses perturbations. Following Clark and Abbott~\cite{clark2024dynamic}, we study the salient aspects of the spectrum of the Jacobian $\mathcal{J}$ (details in App.~\ref{app:jacobian}). The spectrum of $\mathcal{J}$ has $N^2-N$ degenerate eigenvalues $\lambda=-1/p$, which correspond to perturbations belonging entirely to the space of the short-term, plastic couplings $A_{ij}$. The remaining $2N$ eigenvalues are determined by the solutions of the equation
\begin{equation}\label{eq:eigenvalue_equation}
    \det\!\left[
    (\lambda+1/p)\left(\mathcal{J}_{xx}-\lambda \mathbf{I}_N\right)
    +\mathbf{C}
    \right]=0\,, 
\end{equation}
where the $N\times N$ matrix $\mathbf{C}$ contains a contribution coming from the short-term, plastic couplings, and it is given by
\begin{equation}
    \mathbf{C} = \frac{k}{p}\,\Phi'\,(q_{\infty} \mathbf{I}_N + \frac{1}{N}\bphi^*\otimes \bphi^*)\,, 
\end{equation}
and the block of the Jacobian acting in the space of neuron-neuron perturbations, $\mathcal{J}_{xx}$, reads
\begin{equation}
    \mathcal{J}_{xx} = -\mathbf{I}_N + \mathbf{W}^*\Phi'\,.
\end{equation}
Here, $\mathbf{I}_N$ is the $N\times N$ identity matrix, $\bphi^*$ is a $N$-dimensional vector containing the output of the neurons at a fixed point of the neural network dynamics, and $[\Phi']_{ij} = {\phi^*_i}'\delta_{ij}$ is a diagonal matrix whose diagonal elements are the derivatives of the output of the neurons at the fixed point of the dynamics. The quantity $q_{\infty} = \frac{1}{N}\sum_{i=1}^N \langle(\phi^*_i)^2\rangle$ is the self-overlap of the neuron output at the fixed point, and it is equivalent to the long time limit $\lim_{t\to\infty} C_\phi(t,t)$ of the autocorrelation function in Eq.~\eqref{eq:Cphi} from the DMFT. Finally, the matrix $\mathbf{W}^*$ contains the long term and short-term synaptic weights at the fixed point, $[\mathbf{W}^*]_{ij}= J_{ij} + A^*_{ij}$.

The resulting eigenvalue structure, namely the $2N$ coupled
neuron--synapse eigenvalues obtained from Eq.~\eqref{eq:eigenvalue_equation}
together with the $N^2-N$ degenerate eigenvalues $\lambda=-1/p$
coming from $\mathcal{J}_{AA}$, clarifies how short-term synaptic
plasticity affects the nature of excitations around fixed points of the landscape. In the extended system composed by neurons and synapses, the block $\mathcal{J}_{AA}$ always contributes $N^2$ eigenvalues at $-1/p$, corresponding to the intrinsic relaxation of the synapses. When $k=0$, the coupling matrix $\mathbf{C}$ vanishes, so these synaptic modes are completely decoupled from the neural dynamics, and the stability of the neuronal fixed point is determined solely by the spectrum of $\mathcal{J}_{xx}$. For $k>0$, however, the plasticity-induced correction $\mathbf{C}$  provides an additional feedback channel through which activity perturbations influence the neural dynamics. This extra dissipative pathway mirrors the intuition from the analysis of the dynamics in App.~\ref{app:dmft}: synaptic dynamics suppress or soften unstable modes that would otherwise destabilize the purely static network. Finally, in the limit of very large $k$, the fixed points typically lie in a saturated regime of the neuronal nonlinearity, so that $\phi'(x_i^*) \approx 0$. In this case $\Phi' \to \mathbf{0}$ and the effective feedback matrices $\mathcal{J}_{xx} = -\mathbf{I}_N + \mathbf{W}^* \Phi'$ and $\mathbf{C} \propto \Phi'$ reduce to their leak terms, yielding a spectrum of eigenvalues concentrated at $-1$ (neurons) and $-1/p$ (synapses). Thus, linear excitations affecting neurons and synapses become effectively decoupled. 

\begin{figure}
   \includegraphics[width=\columnwidth]{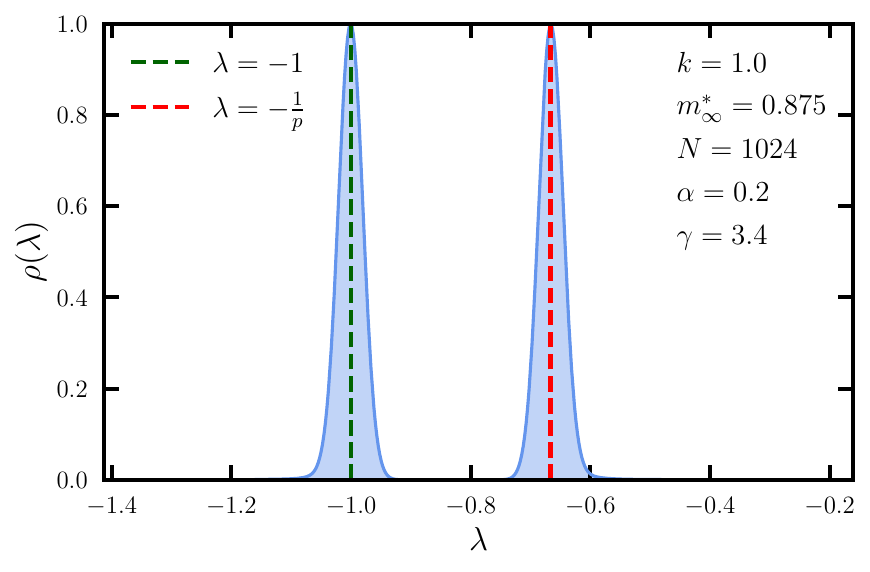}
    \caption{\textbf{Short-term synaptic plasticity and linear excitations.} Eigenvalue spectrum of the Jacobian of the dynamics of the full population of neuron and synapses, at a fixed point of the plastic retrieval regime for $k=1$. The support of the spectrum lies entirely on the negative real axis, and the spectrum is peaked around the theoretical prediction of Sec.~\ref{sec:jacobian} (dashed blue and red lines). 
    }
    \label{fig:jacobian_spectrum}
\end{figure}

In Fig.~\ref{fig:jacobian_spectrum}, we show the spectrum  $\rho(\lambda)$ of of the linearized dynamics of the neural network with short-term synaptic plasticity once it has reached a fixed point in the plastic retrieval phase. 
Even for modest values of the short-term plasticity strength, we see that the eigenvalues of the spectrum are mainly concentrated around the predicted values of $-1$ and $-1/p$, and that the support of $\rho(\lambda)$ lies entirely on the negative side of the real axis. This finding supports the picture brought forth in this work, that short-term synaptic plasticity creates stable basins of attraction in the vicinity of desired memories. 

\subsection{Optimal synaptic timescale}
\begin{figure}[h]
    \includegraphics[width=\columnwidth]{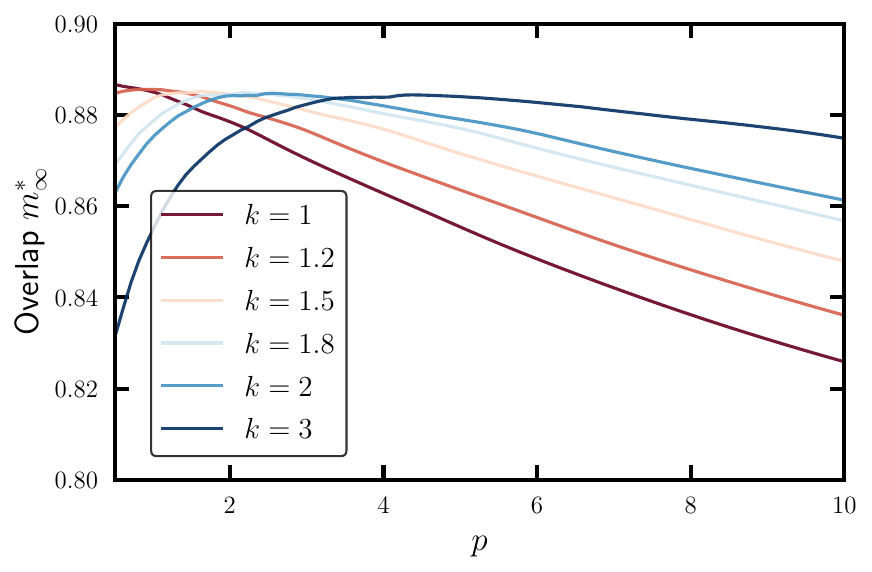} 
    \caption{\textbf{Optimal timescale of plastic retrieval.} Long time overlap $m_\infty^*$ obtained through plastic retrieval at $\alpha=0.2$, $\gamma=3.4$, and at different values of the short-term plastic strength $k$, as a function of the short-term plasticity timescale $p$. The final overlap exhibits a maximum around $p\approx 2$, consistent with the timescale over which transient retrieval takes place.}
    \label{fig:optimal_p}
\end{figure}

In this Section, we discuss the effects of the short-term plasticity timescale $p$ of the plastic dynamics on the retrieval capability of the network. Indeed, from our DMFT analysis in Sec.~\ref{sec:dmft}, we see that short-term synaptic plasticity affects the dynamics by means of a feedback mechanism, whose signal decays over a timescale $p$. If the plastic dynamics is too fast, the synapses $A_{ij}$ track the neural activity too closely, preventing the formation of a stable representation. If the plastic dynamics is too slow, the plastic synapses cannot react in time to stabilize the transient neural dynamics, and catastrophic forgetting is expected to occur. This qualitative analysis suggests the existence of an optimal timescale over which short-term synaptic plasticity should act in order to robustly enlarge the basin of attraction around a desired memory. This is precisely what we demonstrate in Fig.~\ref{fig:optimal_p}. 
For a fixed load $\alpha>\alpha_c$ we solve numerically the DMFT equations \cref{eq:dmft_main_x,eq:m*,eq:eta_eta,eq:Rm,eq:Rphi,eq:Cphi,eq:Kplast} for different values of $p$, and determine the asymptotic overlap of the network with the target pattern  $m^*_\infty$. We then plot the asymptotic value of the overlap, as a function of the short-term plasticity timescale $p$. Different curves are constructed for different values of the short-term plasticity strength $k$. They all display a nonmonotonic behavior as the characteristic short-term plasticity timescale $p$ is varied. All the curves reach a maximum around $p\approx 2$, with the location of the maximum exhibiting only a weak dependence on $k$. A comparison with Fig.~\ref{fig:dmft_vs_sims} (a) shows that the plastic timescale that allows optimal plastic retrieval when $k>0$ roughly matches the time at which the maximum overlap is achieved at $k=0$ during the transient recovery dynamics. This finding corroborates the picture that short-term synaptic plasticity exploits the transient recovery trajectories to lead the neural network toward stable fixed points in the vicinity of the target pattern. Interestingly, we observe that the plastic timescale does not enter at all into the expression of the Lyapunov function $\mathcal{L}$ in Eq.~\eqref{eq:Lyapunov}, which describes the energy landscape in which the network evolves: the metastable states created by short-term synaptic plasticity have thus a complex structure, and the ability of the network to fall into them can be enhanced or suppressed by changing the short-term plasticity timescale. 

\section{Outlook}\label{sec:conclusion}
In this work, we studied how short-term associative synaptic plasticity affects the retrieval capabilities of a recurrent neural network with a long-term Hebbian learning rule. Short-term synaptic plasticity propagates past neural activity to the current state of the network, and it constantly reshapes the effective landscape in which the neurons evolve. The evolution of the neurons in this restricted landscape is thus akin to the motion of a ball on an elastic trampoline, constantly reshaped by the ball's position. By this means, the network is able to recover memories that would be otherwise forgotten, and the network's retrieval abilities are substantially enhanced. We conclude by outlining interesting directions that our work opens the way to.  

First, inspecting the impact of short-term synaptic plasticity on retrieval properties of dense associative memories~\cite{krotov2016dense, lucibello2024exponential} is a relevant question with applications in both biologically plausible learning and artificial neural network. Due to the exponentially large capacity of dense associative memories, any improvement in pattern retrieval capabilities can yield a great impact on memory storage. It is unclear at the moment, and worth investigating, whether the computational workload implied by the update of dense synapses can be circumvented simply by the addition of sparser realizations of short-term synaptic plasticity.

Second, another intriguing direction consists in achieving a fine-grained description of the impact of short-term synaptic plasticity on the basins of attraction of recurrent neural networks. The fact that the dynamics of the system optimize an energy function allows the deployment of techniques from the statistical mechanics of spin-glasses to characterize these metastable states, such as the Franz-Parisi potential~\cite{franz1995recipes} or the Kac-Rice method~\cite{fyodorov2004complexity, ros2019complex}. However, the complex spin-glass structure of the recurrent neural network  poses impressive technical challenges, already without short-term synaptic plasticity, but achieving such an understanding is a very much desirable long-term objective. 

Third, it would be interesting to understand the interplay between short-term synaptic plasticity and noise in the neural network. In biological models  this noise could stem from neural activity in nearby regions, and recent work has shown that adding an out of equilibrium noise can increase significantly the storage capacity of the network~\cite{behera2023enhanced, du2024active}. How does noise affects short-term synaptic plasticity, and how do the qualitative features of this effect change when the ratio between the characteristic timescale of noise and short-term synaptic plasticity are varied is an interesting question to address. 

Finally, a direction worth investigating consists in understanding how short-term synaptic plasticity reshapes the landscape of Hopfield-like recurrent neural network if an external drive is imposed to the system and both short-term plasticity and short-term antiplasticity (the latter can be realized using a negative $k$) are alternatively allowed, before freezing the whole set of synaptic weights of the network, in the spirit of what has been done in~\cite{wakhloo2025associative}. Is it possible to erase one of the long-term stored patterns, and replace it with a new one? If a dynamical stimulus is applied to the network, can limit cycles and complex trajectories be stored within the basin of attraction of a target pattern, or would they destroy the metastable states generated through short-term synaptic plasticity? Progress on these questions should be possible on the basis of the aforementioned study and our work.

\acknowledgments
We thank Itamar D. Landau, Aditya Mahadevan and Atshushi Yamamura for insightful discussions. F.G. acknowledges support from the Stanford Leinweber's Institute of Theoretical Physics. S.G. thanks the Simons foundation and a Schmidt foundation polymath award for support. We thank the Stanford Sherlock cluster for computational resources.

\appendix

\section{Details on Fig.~\ref{fig:fig1_retrieval}(b--d) and Fig.~\ref{fig:fig1_retrieval}(f--h)}\label{app:fig1_details}

In this Appendix, we report the numerical parameters and implementation details used to generate the energy–landscape plots shown in Fig.~\ref{fig:fig1_retrieval} (b--d) and Fig.~\ref{fig:fig1_retrieval} (f--h), where the dynamics of a low-dimensional neural network is simulated, in a parameter regime that allow us to mimic the effects of transient retrieval, catastrophic forgetting, and plastic retrieval, while illustrating the trampoline mechanism induced by short-term synaptic plasticity.  

The system consists of $N=2$ continuous neurons, with
two stored binary patterns: a "spurious" pattern
$\boldsymbol{\xi}^{(1)}=(1,-1)$ and a "target" pattern $\boldsymbol{\xi}^{(2)}=(-1,1)$. The position of the latter is identified by the red cross in Fig.~\ref{fig:fig1_retrieval}.
The fixed Hopfield coupling matrix $\mathbf{J}$ is constructed from these patterns as described in Eq.~(\ref{eq:J_ij}). The system follows the dynamics given by Eq.~\eqref{eq:neuron_dyn} and Eq.~\eqref{eq:stp_dyn}, with the addition of an external field $h \xi_i^{(1)}$ on the right hand side of Eq.~\eqref{eq:neuron_dyn}. The field points along the direction of the spurious pattern, and it is included to destabilize the target memory. Because of this additional magnetic field, the Lyapunov function $\mathcal{L}$ of the 2-neurons system is defined as in Eq.~(\ref{eq:Lyapunov}), with the addition of a term $-h\left(\xi_1^{(1)}\phi_1 + \xi_2^{(1)}\phi_2\right)$.  We use a gain $\gamma=3.4$, and a field strength $h=0.52$. The resulting equations are integrated using a forward Euler scheme with time step $\Delta t=0.01$ for a total of $T=2000$ steps.
The trajectories shown in Fig.~\ref{fig:fig1_retrieval} (b--d, f--h) start from the same initial condition
$\{x_1=0.1,\,x_2=0.4\}$, which leads to the same overlap against both the "spurious" and target patterns. In this way, the "spurious" pattern can significantly interfere with the retrieval process. 
The short-term plasticity strength is taken to be $k=0$ for Fig. \ref{fig:fig1_retrieval} (f--h) and $k=3$, with short-term plastic relaxation time $p=1.0$, for Fig. \ref{fig:fig1_retrieval} (b--d). The three panels in each triptych correspond to times $t\simeq0.01$, $t\simeq1.0$, and $t\simeq4.0$. 

\section{Lyapunov function of RNN with short-term synaptic plasticity}\label{app:Lyapunov}
In this Appendix, we show that the coupled neuron–synapses dynamics admits a Lyapunov function under the assumptions used throughout: the effective connectivity $W_{ij}=J_{ij}+A_{ij}$ is symmetric, the static matrix $J$ is time–independent, and the activation function $\phi$ is strictly increasing (and therefore invertible). These conditions ensure that both the neural subsystem and the synaptic dynamics contribute through dissipative terms to the time derivative of the following Lyapunov function:
\begin{align}
    \begin{split}
        \mathcal{L}
        &\equiv -\frac{1}{2}\sum_{i,j} W_{ij}\,\phi_i\phi_j
        \\&+ \sum_i \int_0^{\phi_i} \phi^{-1}(v)\,\dd v
        + \frac{N}{4k}\sum_{i,j}A_{ij}^2\,,
    \end{split}
\label{eq:L_def}
\end{align}
where $\phi_i=\phi(x_i)$. The three contributions correspond respectively to the synaptic interaction energy, a convex antiderivative term ensuring compatibility with the neuron nonlinearity, and a quadratic potential governing the plastic variables.

We first differentiate the $W$--dependent term.
We obtain
\begin{align}
\begin{split}
\frac{\dd}{\dd t}\!\left( -\frac12 \sum_{i,j} W_{ij}\phi_i\phi_j \right)
= -\sum_i \frac{\dd \phi_i}{\dd t} \sum_j W_{ij}\phi_j\\
  \;-\; \frac12 \sum_{i,j} \frac{\dd A_{ij}}{\dd t}\,\phi_i\phi_j\,,
\end{split}
\end{align}
since $J_{ij}$ is static and only $A_{ij}$ depends on time.
We now proceed with the second term.
Since $\frac{\dd}{\dd u}\phi^{-1}(u)=x$ when $u=\phi(x)$, we obtain
\begin{align}
\frac{\dd}{\dd t}\!\left(\sum_i \int_0^{\phi_i} \phi^{-1}(v)\,\dd v\right)
= \sum_i x_i\,\frac{\dd\phi_i}{\dd t}\,.
\end{align}
Finally,differentiating the synaptic quadratic term:
\begin{align}
\frac{\dd}{\dd t} \left(\frac{N}{4k}\sum_{i,j}A_{ij}^2\right)
= \frac{N}{2k}\sum_{i,j} A_{ij}\frac{\dd A_{ij}}{\dd t}.
\end{align}
The full expression of the time derivative of the Lyapunov function $\mathcal{L}$ becomes
\begin{align}
\begin{split}
    \frac{\dd \mathcal{L}}{\dd t} 
&= \sum_i \frac{\dd\phi_i}{\dd t} \left( x_i - \sum_j W_{ij}\phi_j\right)
  \\&- \frac12 \sum_{i,j} \frac{\dd A_{ij}}{\dd t}\,\phi_i\phi_j +\frac{N}{2k}\sum_{i,j} A_{ij}\frac{\dd A_{ij}}{\dd t}.
\end{split}
\end{align}
From the neuronal dynamics in Eq.~\eqref{eq:neuron_dyn} we obtain
\begin{align}
\sum_i \frac{\dd\phi_i}{\dd t}\Bigl[x_i - \sum_j W_{ij}\phi_j\Bigr]
= -\sum_i \frac{\dd\phi_i}{\dd t}\,\frac{\dd x_i}{\dd t}.
\end{align}
Since $\phi_i=\phi(x_i)$,
\begin{align}
\frac{\dd\phi_i}{\dd t}
= \phi'(x_i)\frac{\dd x_i}{\dd t}
\quad\Longrightarrow\quad
\frac{\dd x_i}{\dd t}
= (\phi^{-1})'(\phi_i)\,\frac{\dd\phi_i}{\dd t}\,,
\end{align}
where we stress that by $(\phi^{-1})'(\phi_i)$ we mean the derivative of the inverse of the function $\phi$, evaluated at $\phi_i$.
Thus
\begin{equation}
-\sum_i \frac{\dd\phi_i}{\dd t}\,\frac{\dd x_i}{\dd t}
= -\sum_i (\phi^{-1})'(\phi_i)\left(\frac{\dd\phi_i}{\dd t}\right)^2 \le 0\,,
\label{eq:Lphi_final}
\end{equation}
since $(\phi^{-1})'(u)>0$ for a strictly increasing activation function. We thus see that the contribution to $\frac{\dd \mathcal{L}}{\dd t}$ from the neuronal dynamics is entirely negative.

We now collect all terms involving $\frac{\dd A_{ij}}{\dd t}$. From above, the $A$-dependent terms are
\begin{align}
 -\frac12 \sum_{i,j} \frac{\dd A_{ij}}{\dd t}\,\phi_i\phi_j
+ \frac{N}{2k}\sum_{i,j}A_{ij}\frac{\dd A_{ij}}{\dd t}.
\label{eq:L_A_raw}
\end{align}
Inserting the dynamical evolution of short-term synaptic plasticity from Eq.~\eqref{eq:stp_dyn} into~\eqref{eq:L_A_raw} and simplifying term-by-term gives
\begin{equation}
 -\frac{N}{2kp}\sum_{i,j}
   \left(A_{ij} - \frac{k}{N}\phi_i\phi_j\right)^2
\le 0\,.
\label{eq:L_A_final}
\end{equation}
This identity follows from completing the square in $A_{ij}$ ad observing that all linear
and quadratic terms cancel exactly.

Summing Eq.~\eqref{eq:Lphi_final} and Eq.~\eqref{eq:L_A_final} together we get
\begin{align}
\frac{\dd\mathcal{L}}{\dd t}
\le 0\,.
\label{eq:L_final}
\end{align}
Both terms are non-positive for $k>0$, so $\mathcal{L}(t)$ is a Lyapunov function.
The bound in Eq.~\eqref{eq:L_final} holds if and only if:
\begin{enumerate}
\item $\dfrac{\dd\phi_i}{\dd t}=0$ for all $i$ (neurons at equilibrium),
\item $A_{ij}=\frac{k}{N}\phi_i\phi_j$ for all $i,j$,
\item which implies $\frac{\dd A_{ij}}{\dd t}=0$ under Eq.~\eqref{eq:stp_dyn}\,,
\end{enumerate}
which concludes our analysis of the Lyapunov function of our recurrent network with short-term synaptic plasticity.

\section{Cavity Calculation of Global Minima}
\label{app:cavity_static}

In this Appendix, we use the static cavity method~\cite{mezard1987, del2014cavity} to obtain a set of self-consistent equations for the fixed point of the recurrent neural network with short-term synaptic plasticity and long-term Hebbian rule. Our starting point are the equation of motion for the action potential of the neurons in the system and the plasticity matrix, given by Eq.~\eqref{eq:neuron_dyn}  and Eq.~\eqref{eq:stp_dyn}. These equations are recapitulated here for completeness, and are given by
\begin{align}
\label{eq:neuron_dyn_pert}
\begin{split}
\partial_{t}x_{i}(t) &= -x_{i}(t)+\sum_{j=1}^{N}W_{ij}(t)\phi(x_{j}(t)) + h_i\\
p\p_t A_{ij}(t) &= -A_{ij}(t) + \frac{k}{N}\phi(x_i(t))\phi(x_j(t))\,, 
\end{split}
\end{align}
where we recall that $W_{ij} \equiv J_{ij} + A_{ij}$, with $J_{ij}$ a Wishart matrix, defined in Eq.~\eqref{eq:J_ij} which stores the long-term memories of the system. Note that in the equation for the neuron's activation, an external field $h_i$ has been added. It will be set equal to zero at the end of the calculation, and it will be used to define the static response functions for the system. 

At the fixed point of the dynamics in Eq.~\eqref{eq:neuron_dyn_pert}, $\partial_t x_i = 0$ and $\partial_t A_{ij} = 0$. The plastic component $A_{ij}$ of the weight matrix reads therefore 
\begin{equation}
A_{ij} = \frac{k}{N} \phi_i \phi_j.
\end{equation}
Thus the fixed-point equation for the post-synaptic potential of neuron $i$ becomes
\begin{align}\label{eq:app_x_i_fixed_point}
    \begin{split}
    x_i &= \sum_{j=1}^N J_{ij} \phi_j
        + \frac{k}{N} \phi_i \sum_{j=1}^N \phi_j^2 + h_i \\
        &= \frac{1}{N}\sum_{\mu=1}^P\sum_{j=1}^N \xi_i^\mu \xi_j^\mu \phi_j
        + \frac{k}{N} \phi_i \sum_{j=1}^N \phi_j^2 + h_i\,.
    \end{split}
\end{align}
We can now introduce a self-overlap for the full network of neurons, $q$, as
\begin{equation}
q \equiv \frac{1}{N} \sum_{j=1}^N \phi_j^2\,,
\end{equation}
and the overlap $m^\mu$ of the output of the network with respect to a given pattern $\xi^\mu$ as
\begin{equation}
    m^\mu \equiv \frac{1}{N} \sum_{i=1}^N \xi_i^\mu \phi_i\,.
\end{equation}
Equation~\eqref{eq:app_x_i_fixed_point} becomes then
\begin{align}
    x_i &= \sum_{\mu=1}^P \xi_i^\mu m^\mu + k q \phi_i + h_i.
\end{align}
The first term on the right hand side contains the influence of the overlap of the network output with the patterns stored in the long-term connectivity matrix $J_{ij}$. Such a contribution is split into a part containing a set of $s\sim O(1)$ condensed patterns, along which we expect the output of the network to align, and a set of $P-s$ uncondensed patterns. We thus obtain
\begin{align}
    x_i
    &= \sum_{\mu^*} \xi_i^{\mu^*} m^{\mu^*}
     + \sum_{\mu>s} \xi_i^\mu m^\mu
     + k q \phi_i + h_i,\,
\end{align}
where the indexes $\mu^*$ run through the indexes of the condensed patterns. Without loss of generality, we have assumed that the condensed patterns are the first $s$ patterns in the connectivity matrix, so that $\mu^*=1,\ldots,s$. 

We add a cavity neuron with index $0$. At the fixed point, the equation for its post-synaptic potential $x_0$ reads:
\begin{align}\label{eq:app_x0}
    x_0 &= \sum_{\mu^*} \xi_0^{\mu^*} \widetilde m^{\mu^*}
        + \sum_{\mu>s} \xi_0^\mu \widetilde m^\mu
        + k \widetilde{q} \phi_0 + h_0\,,
\end{align}
where we denoted by $\tilde{m}^\mu$ and $\tilde{q}$ the order parameters in the system composed by $N+1$ neurons. These quantities are related to the ones in original system composed by $N$ neurons through linear response. Indeed, the addition of a single cavity neuron can be regarded as a small perturbation to the original system. The perturbed, uncondensed overlaps $\widetilde m^\mu$ can therefore be expanded as
\begin{align}
    \begin{split}
        \tilde m^\mu
        &\approx m^\mu
        + \frac{1}{N}\sum_{\nu}\chi^{(m,s)}_{\mu\nu}\,\xi_0^\nu\phi_0\,,
\label{eq:final_tilde_m_app}
    \end{split}
\end{align}
where $\chi^{(m,s)}_{\mu\nu} \equiv \frac{\delta m^\mu}{\delta s^\nu}$ is the response matrix of pattern $\mu$ against a small shift of pattern $\nu$. It can be shown that the change in the self-overlap $\delta q = \widetilde{q} - q$ is $O(1/N)$ and can thus be neglected, as well as the perturbation produced by the cavity neuron to the condensed patterns, since their number is subextensive in the size of the system. Substituting Eq.~\eqref{eq:final_tilde_m_app} back into the fixed point equation Eq.~\eqref{eq:app_x0} we obtain, to leading order in $1/N$,
\begin{align}\label{eq:app_x0_cavity_intermediate}
    \begin{split}
    x_0 &= \sum_{\mu^*} \xi_0^{\mu^*} m^{\mu^*}
    + \sum_{\mu>s} \xi_0^\mu m^\mu\\
        & +\sum_{\mu>s} \xi_0^\mu \left( \frac{1}{N}\sum_{\nu}\chi^{(m,s)}_{\mu\nu}\,\xi_0^\nu\phi_0 \right)
        + k q \phi_0 + h_0.
    \end{split}
\end{align}
In the limit $N \to \infty$, some terms in the equation above concentrate around their average value under different realizations of the patterns $\xi_0^\mu$. The response term becomes thus
\begin{align}\label{eq:app_chi_ms_self_avg}
    \begin{split}
        \sum_{\mu>s} &\xi_0^\mu \left( \frac{1}{N}\sum_{\nu}\chi^{(m,s)}_{\mu\nu}\,\xi_0^\nu\phi_0 \right) \\
        &\approx \left\langle\sum_{\mu>s} \xi_0^\mu \left( \frac{1}{N}\sum_{\nu}\chi^{(m,s)}_{\mu\nu}\,\xi_0^\nu\phi_0 \right) \right\rangle\\
        &= \frac{1}{N}\phi_0 \sum_{\mu,\nu>s} \delta_{\mu\nu} \langle \chi^{(m,s)}_{\mu\nu}\rangle\\
        &= \alpha \phi_0 \frac{1}{P}\sum_{\mu>s} \langle \chi^{(m,s)}_{\mu\mu}\rangle\\
        &\equiv \alpha \overline{\chi}^{(m,s)}\phi_0\,,
    \end{split}
\end{align}
where the last line defines the average longitudinal response of the patterns $\overline{\chi}^{(m,s)}$. Note that in passing from the second to the third line, we neglected the contributions to the susceptibility from the condensed patterns, which is an order $O(1//N)$ contribution. We also used the fact that the patterns are independent, identical random variables with variance one, and that $\sum_{\mu>s} (\xi_0^\mu)^2 \approx P-s \approx \alpha N$. 
Plugging Eq.~\eqref{eq:app_chi_ms_self_avg} 
back into Eq.~\eqref{eq:app_x0_cavity_intermediate} we obtain
\begin{align}\label{eq:app_x0_as_HetaGamma}
    \begin{split}
    x_0 =& \underbrace{\sum_{\mu^*} \xi_0^{\mu^*} m^{\mu^*}}_{H}
        + \underbrace{\sum_{\mu>s} \xi_0^\mu m^\mu}_{\eta}\\
        &+ \underbrace{ (\alpha\,\overline\chi^{(m,s)}
         + k q)}_{\Gamma} \phi_0 + h_0\\
        &\equiv H + \eta + \Gamma \phi_0 + h_0\,,
    \end{split} 
\end{align}
where the second line defines the terms $H$, $\eta$, and $\Gamma$. In the limit of $N\to \infty$, the term $\eta$ becomes a Gaussian noise because of the central limit theorem. Its correlations are given by  
\begin{align}
    \begin{split}
        \langle \eta^2 \rangle&= \sum_{\mu,\nu>s} \langle\xi_0^\mu \xi_0^\nu m^\mu m^\nu\rangle \\
        &= \sum_{\mu>s} \langle[m^\mu]^2\rangle \\
        &= N\frac{P}{N}\frac{1}{P}\sum_{\mu>s} \langle[m^\mu]^2\rangle \\
        &= \alpha N \left[\frac{1}{P}\sum_{\mu>s} \langle[m^\mu]^2\rangle\right] \,,
    \end{split}
\end{align} 
We can thus interpret the term $\eta$ as noise stemming from the uncondensed patterns along which the system does not align. The term $\Gamma$ is a screening term that renormalizes, by means of the response function and the short-term synaptic plasticity, the output of the cavity neuron back onto the neuron itself. To solve for the unknown response term and noise-noise correlations, we perturb again the original $N$-neurons system with an additional uncondensed, cavity pattern $\xi^0$ with a source $s^0$. The addition of a pattern acts again as a small perturbation to the original $N$-neurons system. The cavity overlap $m^0$ can be thus written using linear response, as
\begin{align}
    \begin{split}
        m^0 &= \frac{1}{N}\sum_{i=1}^N \xi_i^0 \phi_i + \frac{1}{N}\sum_{i,j}^N \xi_i^0 \phi'_i \chi^{(x, h)}_{ij} \xi_j^0 m^0 + s^0 \\
        &= \frac{1}{N}\sum_{i=1}^N \xi_i^0 \phi_i + \frac{1}{N}\sum_{i,j}^N \xi_i^0 \chi^{(\phi, h)}_{ij} \xi_j^0 m^0 + s^0\\
        &\approx \eta^0  + \overline{\chi}^{(\phi,h)} m^0 + s^0\,,
    \end{split}
\end{align}
where in the first line we introduced the response $\chi^{(x,h)}_{ij} \equiv \frac{\delta x_i}{\delta h_j}$ of the potential of neuron $i$ to a small external field applied to neuron $j$. This response is related to the response matrix $\chi^{(\phi,h)} \equiv \frac{\delta \phi_i}{\delta h_j}$, which encode the response of the output of neuron $i$ against a small external field applied to neuron $j$. The relation between these two response matrices is $\chi_{ij}^{(\phi,h)} = \chi_{ij}^{(x,h)}\phi'_i$. Finally, in the last line, we used the fact that in the thermodynamic limit the response term self-averages around its mean value, $\overline{\chi}^{(\phi,h)} \equiv \frac{1}{N} \sum_i \chi_{ii}^{(\phi,h)}$. Solving for $m^0$ yields
\begin{equation}\label{eq:app_cavity_m0}
    m^0 = \frac{\eta^0 + s^0}{1 - \overline{\chi}^{(\phi,h)}}\,.
\end{equation}
By the central limit theorem, the term $\eta^0$ becomes a Gaussian noise with zero mean and variance
\begin{equation}
    \langle [\eta^0]^2 \rangle = \frac{1}{N^2} \sum_{ij} \langle \xi_i^0 \xi_j^0 \phi_i \phi_j \rangle = \frac{1}{N} q\,.
\end{equation}
Since $m^0$ is proportional to $\eta^0$, we conclude that $m^0$ is also a Gaussian noise, with zero mean and variance
\begin{equation}
    \langle (m^0)^2 \rangle
    = \frac{\langle (\eta^0)^2 \rangle}{(1 - \overline\chi^{(\phi,h)})^2} = \frac{q}{N(1 - \overline\chi^{(\phi,h)})^2}\,.
\end{equation}
The variance of the noise $\eta$ in the cavity equation for $x_0$, Eq.~\eqref{eq:app_x0_as_HetaGamma}, is thus
\begin{equation}
    \sigma^2 \equiv \langle \eta^2 \rangle
    = \alpha N \langle (m^0)^2 \rangle
    = \frac{\alpha q}{(1 - \overline\chi^{(\phi,h)})^2}.
\end{equation}
Moreover, the average susceptibility of an uncondensed pattern $\overline\chi^{(m,s)} = \delta m^0 / \delta s^0$ is related to the average susceptibility of a neuron output. By taking the derivative of Eq.~\eqref{eq:app_cavity_m0} with respect to $s^0$, we obtain 
\begin{equation}\label{eq:app_relation_chim_chiphi}
    \overline\chi^{(m,s)} = \frac{1}{1 - \overline\chi^{(\phi,h)}}\,.
\end{equation}
we thus need only one last self-consistent equation for the susceptibility of a neuron output. It is convenient at this point to introduce the function $g(x) \equiv \tanh \gamma x$, to distinguish the application of a nonlinear function to the neuron output $\phi_0$. We then apply $g$ to both sides of Eq.~\eqref{eq:app_x0_as_HetaGamma}, obtaining a self-consistent equation for the output $\phi_0$, namely
\begin{equation}\label{eq:app_cavity_phi0}
    \phi_0 = g\left(H + \eta + \Gamma\phi_0 + h_0\right)\,.
\end{equation}
We can then obtain an expression for the average output susceptibility $\overline{\chi}^{(\phi,h)}$ by observing that $\frac{\delta \phi_0}{\delta \eta} =  \frac{\delta \phi_0}{\delta h_0}$. Then, using integration by parts, we obtain 
\begin{align}\label{eq:self_consistent_chi}
    \begin{split}
        \overline{\chi}^{(\phi,h)} &=  \left\langle \frac{\delta \phi_0}{\delta \eta} \right\rangle_{\eta,\xi^{\mu^*}}
        = \frac{1}{\sigma^2} \left\langle \eta \phi_0 \right\rangle_{\eta,\xi^{\mu^*}} \\
        &= \frac{\left(1 - \overline{\chi}^{(\phi,h)}\right)^2}{\alpha q} \left\langle \eta \phi_0 \right\rangle_{\eta,\xi^{\mu^*}}\,.
    \end{split}
\end{align}
Putting everything together, we can write all the self-consistent equations for the fixed point of the network dynamics. The index $0$ can be dropped, since all the neurons in the fully connected system are equivalent. The self-consistent set of equations read 
\begin{align}\label{eq:cavity_phi} 
    \begin{split}
        \phi &= g\!\left(
            \sum_{\mu^*} \xi^{\mu^*} m^{\mu^*}
            + \eta
            + \Gamma \phi + h
        \right)\,,\\
        \langle \eta^2\rangle &\equiv \sigma^2 = \frac{\alpha q}{(1 - \overline{\chi}^{(\phi,h)})^2}\,,\\
        q &= \langle \phi^2 \rangle_{\eta,\xi^{\mu^*}}\,,\\
        \Gamma &= \frac{\alpha}{1 - \overline\chi^{(\phi,h)}}  + k q\,, \\
        \overline\chi^{(\phi,h)}
        &= \frac{(1 - \overline\chi^{(\phi,h)})^2}{\alpha q}\,
           \langle \eta\,\phi \rangle_{\eta,\xi^{\mu^*}}\,, \\
        m^{\mu^*} &= \langle \xi^{\mu^*} \phi \rangle_{\eta,\xi^{\mu^*}} 
    \end{split}
\end{align}
where we recall that $\langle \cdot \rangle_{\eta,\xi^{\mu^*}}$ denotes the average over the Gaussian noise $\eta \sim \mathcal{N}(0,\sigma)$ and the realization of mean field condensed patterns $\xi^{\mu^*}$. Eq.~\eqref{eq:cavity_phi} can be rewritten by redefining the noise as $\eta=\sigma z$, with $z$ a Gaussian random variable of mean zero and unit variance. The final result is
\begin{align}\label{eq:app_cavity_phi_with_z} 
    \begin{split}
        \phi &= g\!\left(
            \sum_{\mu^*} \xi^{\mu^*} m^{\mu^*}
            + \sigma z
            + \Gamma \phi + h
        \right)\,,\\
       \sigma^2 &= \frac{\alpha q}{(1 - \overline{\chi}^{(\phi,h)})^2}\,,\\
        q &= \langle \phi^2 \rangle_{z,\xi^{\mu^*}}\,,\\
        \Gamma &= \frac{\alpha}{1 - \overline\chi^{(\phi,h)}}  + k q\,, \\
        \overline\chi^{(\phi,h)}
        &= \frac{1 - \overline\chi^{(\phi,h)}}{\sqrt{\alpha q}}\,
           \left\langle \frac{\dd \phi}{\dd z}\right\rangle_{z, \xi^{\mu^*}}\,, \\
        m^{\mu^*} &= \langle \xi^{\mu^*} \phi \rangle_{z, \xi^{\mu^*}}\,,
    \end{split}
\end{align}
where $\langle\cdot\rangle_{z,\xi^{\mu^*}}$ denotes an average over the realizations of the Gaussian noise $z$ and the condensed pattern $\xi^{\mu^*}$, and we used integration by part to rewrite the expression of $\overline{\chi}^{(\phi,h)}$. By considering the case  of a single condensed overlap $m^*$, Eq.~\eqref{eq:app_cavity_phi_with_z} corresponds to \cref{eq:cavity_overlap,eq:cavity_phi_main,eq:cavity_phi_main,eq:sigma2_Gamma,eq:q,eq:cavity_chi} of the main text.

\subsection{Breakdown of the cavity and spin-glass transition}\label{app:cavity_breakdown}
The fixed point described by the cavity equations, \cref{eq:cavity_overlap,eq:cavity_phi_main,eq:cavity_phi_main,eq:sigma2_Gamma,eq:q,eq:cavity_chi} is not guaranteed to be stable. In particular, the fixed point is prone to instabilities against small random perturbations of the neurons' post-synaptic potentials. To this end, we add to each neuron a weak Gaussian random field $\delta\zeta_i$, with zero mean and unit variance, and study the induced shift of the fixed point.

The perturbation $\delta\zeta_i$ first alters the neuronal activities, then propagates to the uncondensed pattern overlaps, and finally feeds back to the neurons. Linearizing around the fixed point, the resulting displacement of $x_i$ can be written as
\begin{equation}
    \delta x_i
    = \frac{1}{N}\sum_{\mu,\nu=1}^P \sum_{j,k=1}^N
      \chi^{(m,s)}_{\mu\nu}\,\xi_i^\mu \xi_j^\nu\,
      \chi^{(\phi,h)}_{jk}\,\delta x_k
      + \delta\zeta_i\,,
\end{equation}
where $\chi^{(m,s)}_{\mu\nu} \equiv \frac{\delta m^\mu}{\delta s^\nu}$ is the static susceptibility of the uncondensed overlaps to a small source $s^\nu$, and
$\chi^{(\phi,h)}_{jk} \equiv \frac{\delta\phi_j}{\delta h_k}$ is the static susceptibility of the neuronal outputs to small external fields. Both type of response simultaneously appear due to a screening effect: perturbations to the neurons output propagates to the uncondensed overlaps, which in turn backpropagates to the neurons themselves.

In the thermodynamic limit, both response matrices self-average and become effectively diagonal on the relevant scales. By defining the mean squared susceptibilities $\overline\chi^{(m,s)}_2$, $\overline\chi^{(\phi,h)}_2$ as 
\begin{align}
    \begin{split}
        \overline\chi^{(m,s)}_2 &\equiv \frac{1}{P}\sum_{\mu=1}^P \left\langle\left[\chi^{(m,s)}_{\mu\mu}\right]^2\right\rangle\\
        \overline\chi^{(\phi,h)}_2 &\equiv \frac{1}{N}\sum_{i=1}^N \left\langle\left[\chi^{(\phi,h)}_{ii}\right]^2\right\rangle\,,
    \end{split}
\end{align}
and exploiting the fact that the patterns $\xi_i^\mu$ are independent with unit variance, the average squared displacement of $x_i$ under the random field $\delta\zeta_i$ can be written as
\begin{equation}
    \mathbb{E}[\delta x_i^2]
    = \frac{1}{1 - \alpha\,\overline{\chi}^{(m,s)}_2\overline\chi^{(\phi,h)}_2}
    \equiv \Xi\,,
\end{equation}
where $\mathbb{E}[\cdot]$ denotes the average over the Gaussian perturbations $\delta\zeta_i$, and $\Xi$ defines the \emph{fragility} of the fixed point.

The fragility $\Xi$ quantifies the nonlinear susceptibility of the network to random field perturbations. When $\Xi$ diverges, small random perturbations produce macroscopic rearrangements of the fixed point, and the system undergoes a spin glass transition. At this point, the single-site cavity description based on a unique stable fixed point breaks down. The fixed point cavity equations, \cref{eq:cavity_overlap,eq:cavity_phi_main,eq:cavity_phi_main,eq:sigma2_Gamma,eq:q,eq:cavity_chi}, are therefore valid as long as 
\begin{equation}\label{eq:replicon_chi2}
    1 - \alpha\,\overline\chi^{(m,s)}_2\overline\chi^{(\phi,h)}_2 \geq 0\,.
\end{equation}
Expression for the mean squared susceptibilities $\overline\chi^{(m,s)}_2$, $\overline\chi^{(\phi,h)}_2$ can be derived through a similar procedure to the one followed in Sec.~\ref{app:cavity_static} for the derivation of the susceptibilities. The final result is
\begin{align}\label{eq:app_chi2}
    \begin{split}
        \overline{\chi}^{(\phi,h)}_2 &= \frac{\left(1-\overline{\chi}^{(\phi,h)}\right)^2}{\alpha q} \left\langle\left[\frac{\dd \phi}{\dd z}\right]^2\right\rangle_{z,\xi^{\mu^*}}\\
        \overline{\chi}_2^{(m,s)} &= \frac{1}{\left(1 - \overline{\chi}^{(\phi,h)} \right)^2}\,.
    \end{split}
\end{align}
Plugging Eq.~\eqref{eq:app_chi2} into Eq.~\eqref{eq:replicon_chi2}, we obtain the stability condition
\begin{equation}
    \left\langle\left[\frac{\dd \phi}{\dd z}\right]^2\right\rangle_{z,\xi^{\mu^*}} \leq q\,,
\end{equation}
which corresponds to Eq.~\eqref{eq:cavity_braking} of the main text, when we specialize to the case of a single condensed pattern.

\subsection{Phase boundaries with short-term synaptic plasticity}\label{app:boundary_para}

In this Section, we determine the critical value of the gain at zero network load, $\gamma_c(k)$, that separates retrieval phase from the paramagnetic phase, and the boundary in the $(\gamma,\alpha)$ plane that separates the paramagnetic phase from the spin-glass phase, $\gamma_g(\alpha,k)$. 

To derive $\gamma_c$, we consider the fixed point of the recurrent neural network in the presence of a single condensed pattern, \cref{eq:cavity_overlap,eq:cavity_phi_main,eq:cavity_phi_main,eq:sigma2_Gamma,eq:q,eq:cavity_chi,eq:cavity_braking}, in the case where the retrieval is zero, $m^{*} = 0$, and the number of patterns loaded in the network is subextensive in the network size, $\alpha=0$.  Then, the fixed point equation for the output of the cavity neuron, Eq.~\eqref{eq:cavity_phi_main}, reads
\begin{equation}\label{eq:phi_paramagnet_alpha0}
    \phi = \tanh (\gamma k q \phi)\,,
\end{equation}
with the self-overlap $q$ given by Eq.~\eqref{eq:q} in the main text. The stable solution to this equation is $\phi=0$. We now apply a small perturbation $\delta m^*$ to the condensed overlap. This perturbation  shifts the argument of the hyperbolic tangent in Eq.~\eqref{eq:phi_paramagnet_alpha0} by an amount $\xi^*\delta m^*$, which in turn produces a shift $\delta \phi$ to the output of the cavity neuron at the fixed point. The equation satisfied by $\delta \phi$ is obtained by expanding Eq.~\eqref{eq:phi_paramagnet_alpha0}, and it reads
\begin{equation}
    \delta \phi = \gamma \left[ \xi^* \delta m^*\right]\,.
\end{equation}
Multiplying both sides of this equation by the condensed pattern $\xi^*$, and averaging over the realizations of $\xi^*$, we obtain an equation for $\delta m^*$, which reads
\begin{equation}\label{eq:app_m*_perturbing_paramagnet}
    \delta m^* = \gamma \delta m^*\,.
\end{equation}
Equation~\eqref{eq:app_m*_perturbing_paramagnet} yields a self-consistent condition to determine the magnitude of the overlap perturbation $\delta m^*$. The paramagnetic phase is characterized by the trivial solution $\delta m^*$. However, we observe that there is a critical value of the gain, $\gamma_c$, such that nonzero solutions are possible. $\gamma_c$ is obtained from Eq.~\eqref{eq:app_m*_perturbing_paramagnet}, and it reads
\begin{equation}
    \gamma_c^{-1}(k) = 1\,.
\end{equation}
This equation determines the transition, at $\alpha=0$, from the paramagnetic to the retrieval phase. Since the plastic feedback is mediated by the self-overlap $q$, the boundary between the paramagnetic and the retrieval phase is left unchanged with respect to the case $k=0$. This is the result reported in Eq.~\eqref{eq:gamma_c_gamma_g} of the main text.

We now address how short-term synaptic plasticity affects the shape of the boundary $\gamma_g(\alpha)$ between the paramagnetic phase and the spin glass phase. This transition corresponds to the point where the single-site cavity description breaks down due to the divergence of the fragility $\Xi$, as discussed in Appendix~\ref{app:cavity_breakdown}. The condition for this breakdown is given by Eq.~\eqref{eq:replicon_chi2}, when the inequality is saturated. In the paramagnetic phase ($m^*= q=0$), the membrane potential of a cavity neuron is $x = \Gamma \phi$ (see \cref{eq:cavity_overlap,eq:cavity_phi_main,eq:cavity_phi_main,eq:sigma2_Gamma,eq:q,eq:cavity_chi,eq:cavity_braking}). Since the membrane potential is independent from the realizations of the Gaussian field $z$, we have that $\left[\overline{\chi}^{(\phi,h)}\right]^2 = \overline{\chi}_2^{(\phi,h)}$. Therefore, the instability condition in Eq.~\eqref{eq:replicon_chi2} becomes
\begin{equation}\label{eq:app_instability_chi}
    1-\alpha \frac{\left[\overline{\chi}^{(\phi,h)}\right]^2}{\left[1 -\overline{\chi}^{(\phi,h)}\right]^2} = 0
\end{equation}
At the spin-glass transition, this equation yields an expression for the linear output susceptibility $\overline{\chi}^{(\phi,h)}$ in terms of the network load $\alpha$, namely, 
\begin{equation}\label{eq:app_chi_m1}
    \left[\overline{\chi}^{(\phi,h)} \right]^{-1} = 1 +\sqrt{\alpha}\,.
\end{equation}
On the other hand, taking the derivative with respect to an external field $h$ of the paramagnetic,  fixed point equation $\phi=g(\Gamma\phi + h)$ yields 
\begin{equation}\label{eq:app_chi_para}
    \overline{\chi}^{(\phi,h)} = \frac{\gamma}{1 - \gamma\Gamma}\,.
\end{equation}
Combining together Eq.~\eqref{eq:app_chi_m1} and Eq.~\eqref{eq:app_chi_para} we obtain an expression for the output gain $\gamma_g(\alpha)$
at the transition between the paramagnetic and spin-glass phase, which reads
\begin{align}
    \begin{split}
        \gamma_g(\alpha,k)^{-1} &= \left[ \overline{\chi}^{(\phi,h)} \right]^{-1} + \Gamma \\
        &= 1 + \sqrt{\alpha} + \frac{\alpha}{1 - \overline{\chi}^{(\phi,h)}} \\
        &= 1 + \sqrt{\alpha} + \frac{\sqrt{\alpha}}{\overline{\chi}^{(\phi,h)}} \\
        &= (1 + \sqrt{\alpha})^2\,,
    \end{split}
\end{align} 
 where in the second line we used the definition of $\Gamma$ in Eq.~\eqref{eq:sigma2_Gamma} with $q=0$, and Eq.~\eqref{eq:app_chi_m1}, in the third line we used the instability condition in Eq.~\eqref{eq:app_instability_chi}, and in the fourth line we used again Eq.~\eqref{eq:app_chi_m1}. We have thus obtained the result displayed on the second line of Eq.~\eqref{eq:gamma_c_gamma_g} of the main text.

\section{DMFT Derivation}
\label{app:dmft}

In this Appendix, we discuss how to derive the dynamical mean field equations for our recurrent neural network with Hebbian rule and short-term synaptic plasticity. 
The equation for the time evolution of the plastic weights $A_{ij}$ can be obtained by integration of Eq.~\eqref{eq:stp_dyn}, obtaining
\begin{equation}
A_{ij}(t)=e^{-t/p}A_{ij}(0)+\frac{k}{pN}\int_{0}^{t}e^{-(t-\tau)/p}\,\phi_i(\tau)\phi_j(\tau)\,\dd\tau\,. \label{eq:A-sol_app_dmft}
\end{equation}
We now define the kernel $K_p(\Delta)=\frac{1}{p}e^{-\Delta/p}\Theta(\Delta)$, with $\Theta(x)$ the Heaviside theta function, and autocorrelation function $C_\phi(t,\tau)$ for the neuron outputs as 
\begin{equation}
C^\phi(t,\tau)=\frac{1}{N}\sum_{j=1}^{N}\phi_j(t)\phi_j(\tau)\,. \label{eq:Cphi_app_dmft}
\end{equation}
Then the contribution coming from the plastic weights to the dynamics of the neuron outputs in Eq.~\eqref{eq:neuron_dyn} is
\begin{align}
    \begin{split}
        \sum_{j}A_{ij}(t)\phi_j(t)&=e^{-t/p}\sum_j A_{ij}(0)\phi_j(t) \\
        & + k\int_{0}^{t}K_p(t-\tau)\,C_\phi(t,\tau)\,\phi_i(\tau)\,\dd\tau. \label{eq:Aphi_app_dmft}
    \end{split}
\end{align}
since $A_{ij}(0)=0$, only the non-local self-interaction term survives, similarly to the results obtained by~\cite{clark2024dynamic}. Inserting Eq.~\eqref{eq:Aphi_app_dmft} into Eq.~\eqref{eq:neuron_dyn}, we obtain the following expression: 
\begin{align}
\partial_t x_i(t)
&= -x_i(t) + \sum_{j=1}^{N}J_{ij}\phi_j(t) \nonumber \\
&+ k\int_{0}^{t}K_p(t-\tau)\,C_\phi(t,\tau)\,\phi_i(\tau)\,\dd\tau. 
\end{align}
Since we already dealt with the synaptic term, this can be left out in the following computations. We will focus on the term with $J_{ij}$ only, and put back the synaptic kernel in the end. As done in App.~\ref{app:cavity_static}, we introduce a dynamical overlap for each pattern $\mu=1,\ldots,P$,
\begin{equation}
    m^\mu(t) = \frac{1}{N} \sum_{i=1}^N \xi_i^\mu \phi_i(t)\,,
\end{equation}
and we split the set of patterns $\xi_i^\mu$ into a set of condensed patterns, denoted without loss of generality by $\{\mu^\ast\}=\{1,\dots,s\}$,  and a set of uncondensed patterns, for $\mu>s$. The overlap against condensed patterns are of order $m^{\mu^{*}}(t) \sim O(1)$, while the overlaps against the uncondensed patterns are of order $m^{\mu}(t) \sim O(N^{-1/2})$. We are going to add in two distinct steps two distinct cavity variables (a neuron and a pattern respectively, as done in App.~\ref{app:cavity_static}) to the dynamical evolution of a neuron in the absence of short-term synaptic plasticity: 
\begin{align}\label{eq:app_ptxi_k0}
\partial_tx_i(t) = -x_i(t) + \sum_{\mu*}\xi_i^{\mu*}m^{\mu*}(t) + \sum_{\mu>s}\xi_i^{\mu}m^{\mu}(t)
\end{align}
We first consider adding an extra neuron $x_0$, with an extra pattern component $\xi_0^{\mu} \quad \forall \quad \mu$. We denote with the symbol $\widetilde\cdots$ the variable in the $N+1$-neuron system, where the cavity has been inserted. The addition of a cavity variable perturbs the original $N$-neurons system, but the perturbation is small. Indeed the perturbed overlaps of the system with uncondensed patterns, $\widetilde m^\mu$, can be written as  
\begin{align}
    \widetilde{m}^{\mu}(t)=\frac{1}{N}\sum_{i=1}^{N}\xi_{i}^{\mu}\widetilde{\phi}_{i}(t) + \underbrace{\frac{1}{N}\xi_0^{\mu}\phi_0(t)}_\text{$h^{\mu}(t)$}\,,
\end{align}
from which we see that the perturbation induced by the cavity neuron is of order $O(1/N)$. We thus expand the perturbed uncondensed overlaps using linear response,
\begin{align}
    \begin{split}
        \widetilde{m}^{\mu}(t) = &  m^{\mu}(t)  
        + \sum_{\nu}\int_{0}^t \dd\tau \frac{\delta m^{\mu}(t)}{\delta h^{\nu}(\tau)}h^{\nu}(\tau) \\
        =& m^{\mu}(t) + \frac{1}{N}\sum_{\nu} \int_0^t\dd\tau \frac{\delta m^{\mu}(t)}{\delta h^{\nu}(\tau)}\xi_0^{\nu}\phi_0(\tau)\,,
    \end{split}
\end{align}
where $\frac{\delta m^\mu(t)}{\delta h(\tau)}$ is the dynamic response function of an uncondensed pattern at time $t$ when the overlap pattern $m^\nu(t)$ is instantaneously shifted by a kick of size $\delta h(\tau)$ at time $\tau$. Using linear response, the equation of motion of the cavity neuron $x_0$ reads 
\begin{align}
\label{dmft_first_cavity}
    \begin{split}
        \p_t x_0(t) &= -x_0(t) + \sum_{\mu^*}\xi_0^{\mu*} \widetilde m^{\mu^*}(t) + \sum_{\mu>s} \xi_0^{\mu} \widetilde m^\mu(t) \\
        &\approx -x_0(t) + \sum_{\mu^*}\xi_0^{\mu*}m^{\mu*}(t) + \underbrace{\sum_{\mu>s}\xi_0^{\mu}m^{\mu}(t)}_{\eta_0(t)}\\
        & + \int_0^t\dd\tau \frac{1}{N}\sum_{\nu > s, \mu>s} \frac{\delta m^{\mu}(t)}{\delta h^{\nu}(\tau)}\xi_0^{\nu}\xi_0^{\mu}\phi_0(\tau)\,.
    \end{split}
\end{align}
Note that the perturbation from the cavity to the condensed pattern is subleading with respect to the other contributions and can thus be neglected. The term $\sum_{\mu>s} \xi_0^\mu m^\mu$ is independent from the dynamics of the cavity neuron, and is of order $O(1)$. Let us redefine it as
\begin{align}
    \eta_0(t) \equiv \sum_{\mu>s}\xi_0^{\mu}m^{\mu}(t)\,.
\end{align}
By the central limit theorem, in the thermodynamics limit the statistics of $\eta_0(t)$ converges to the statistics of a colored Gaussian noise with zero mean, 
$\langle \eta_0(t)\rangle = 0$, and autocorrelation function given by
\begin{align}
    \begin{split}
        \langle \eta_0(t) \eta_0(t')\rangle &=  \sum_{\mu>s, \nu>s}\langle\xi_0^{\mu}\xi_0^{\nu}m^{\mu}(t)m^{\nu}(t')\rangle \\
        &=\sum_{\mu>s, \nu>s}\langle\xi_0^{\mu}\xi_0^{\nu} \rangle \langle m^{\mu}(t)m^{\nu}(t')\rangle\\
        &= N\alpha \left[\frac{1}{P}\sum_{\mu>s}\langle m^{\mu}(t)m^{\mu}(t') \rangle\right]  \\
        &\equiv \alpha N C_m(t, t')\,, 
    \end{split}
\end{align}
where the last line defines the uncondensed pattern autocorrelation function $C_m(t,t')$. Moreover, the response term concentrates around its mean value, yielding
\begin{align}
    \begin{split}
        \frac{1}{N}&\sum_{\nu>s, \mu>s}\frac{\delta  m^{\mu}(t)}{\delta h^{\nu}(\tau)} \xi_0^{\mu}\xi_0^{\nu} \\
        &\approx \frac{1}{N}\sum_{\nu>s, \mu>s} \left\langle \frac{\delta m^{\mu}(t)}{\delta h^{\nu}(\tau)}\xi_0^{\mu}\xi_0^{\nu}\right\rangle \\
        &= \frac{1}{N}\sum_{\mu>s} \left\langle\frac{\delta m^{\mu}(t)}{\delta h^{\mu}(\tau)}\right\rangle  \\
        &=\alpha  \left[\frac{1}{P}\sum_{\mu>s}\left\langle\frac{\delta m^{\mu}(t)}{\delta h^{\mu}(\tau)}\right\rangle\right]
    \equiv  \alpha R_m(t, \tau)\,,
    \end{split}
\end{align}
where the last line defined the uncondensed pattern response function $R_m(t,\tau)$. Putting everything together back into Eq.~\eqref{dmft_first_cavity}, we obtain an effective equation for the dynamical evolution of the cavity neuron $x_0(t)$ as
\begin{align}\label{eq:app_dmft_x0}
    \begin{split}
        \partial_tx_0(t) &= -x_0(t) + \sum_{\mu*} \xi_0^{\mu*}m^{\mu*}(t) \\
        &+ \alpha\int_0^{t} \dd\tau R_{m}(t, \tau) \phi_0(\tau) + \eta_0(t) \\
        \langle \eta_0(t)\rangle& = 0, \quad \langle \eta_0(t)\eta_0(t')\rangle \equiv \alpha N C_m(t, t')\,.
    \end{split}
\end{align}
To proceed further, we need to describe how do the overlaps with uncondensed patterns evolve.

We now considered adding to the original $N$-neurons system one uncondensed, cavity pattern $\xi^0_i$. The perturbed system now consists of $N$ neurons and $P+1$ patterns. As usual, we denote by $\widetilde\cdots$ the variables in the $P+1$-patterns system, as perturbed by the additional cavity variable. Taking advantage of the fact that the addition of a cavity pattern is a small perturbation to the dynamics of the original $P$-pattern system in the thermodynamic limit, we use linear response theory to rewrite the cavity overlap  $m^0(t)$ as
\begin{align}\label{eq:app_cavity_dyn_m0_initial}
    \begin{split}
       m^{0}(t) &\equiv \frac{1}{N}\sum_{i=1}^{N} \xi_i^0 \tilde{\phi_i}(t) \\
        &= \underbrace{\frac{1}{N}\sum_{i=1}^N \xi_i^0\phi_i(t)}_{\eta^0(t)} \\
        &+ \int_0^t \dd\tau \frac{1}{N}\sum_{i,j}\frac{\delta \phi_i(t)}{\delta h_j(\tau)}\xi_i^0\xi_j^0 m^0(\tau)\,,
    \end{split}
\end{align}
where $\frac{\delta \phi_i(t)}{\delta h_j(\tau)}$ is the dynamical response of the output of neuron $i$ at time $t$ when neuron $j$ has been perturbed by a small instantaneous field of intensity $\delta h$ at time $\tau$.  In the thermodynamic limit, the statistics of the term $\eta^0(t)$ approaches the one of a Gaussian noise with mean $\langle \eta^0(t)\rangle=0$ and autocorrelations
\begin{align}
    \begin{split}
        \langle \eta^0(t)\eta^0(t')\rangle &=  \frac{1}{N^2}\sum_{i,j}\langle\xi_i^0\xi_j^0\phi_i(t)\phi_j(t') \rangle \\
        &= \frac{1}{N^2}\sum_i \langle \phi_i(t)\phi_i(t') \rangle \\
        &\equiv  \frac{1}{N}C_{\phi}(t,t')\,,
    \end{split}
\end{align}
where the last line defines the autocorrelation function of the output of the population of neurons in the original, unperturbed system, $C_\phi(t,t')$. On the other hand, in the thermodynamics limit, the linear response term concentrates around its average value
\begin{align}
    \begin{split}
        \frac{1}{N}\sum_{i,j} \xi_i^0 \xi_j^0  \frac{\delta \phi_i(t)}{\delta h_j(\tau)} &=\frac{1}{N}\sum_{i,j}\langle \xi_i^0 \xi_j^0\rangle \left\langle \frac{\delta \phi_i(t)}{\delta h_j(\tau)}\right\rangle \\
        &=
        \frac{1}{N}\sum_i\left\langle \frac{\delta \phi_i(t)}{\delta h_i(\tau)}\right\rangle  \\
        &\equiv R_{\phi}(t, \tau)\,,
    \end{split}
\end{align}
where the last line defines the response function of the output of a neuron at time $t$, when the neuron has been perturbed by a point kick at time $\tau$. Putting everything together back into the equation for the cavity uncondensed overlap, Eq.~\eqref{eq:app_cavity_dyn_m0_initial}, and adding a small external auxiliary field $h^0(t)$, which will be used for the computation of the response functions, we obtain an effective equation for the cavity uncondensed overlap, namely,
\begin{equation}\label{eq:app_cavity_m0_final}
m^{0}(t)=\eta^{0}(t)+\int_{0}^{t}\dd\tau R_{\phi}(t,\tau)m^{0}(\tau)+h^{0}(t)\,.
\end{equation}
we can use this equation to relate the response of neuron outputs to the response of uncondensed patterns. Taking a functional derivative of both sides of Eq.~\eqref{eq:app_cavity_m0_final} with respect to the field $h^0(\tau)$, and averaging over the realization of the noise $\eta(t)$ and $\eta^0(t)$
\begin{align}\label{eq:app_volterra_RmRphi}
    \begin{split}
        R_m(t,t') &=\left\langle\frac{\delta m^{0}(t)}{\delta h^0(t^{\prime})}\right\rangle \\
        &= \delta(t-t^{\prime})+\int_{0}^{t}\dd\tau R_{\phi}(t,\tau)\left\langle\frac{\delta m^{0}(\tau)}{\delta h^{0}(t^{\prime})}\right\rangle\\
        &=\delta(t-t^{\prime})+\int_{t^{\prime}}^{t}\dd\tau R_{\phi}(t,\tau)R_{m}(\tau,t^{\prime})
    \end{split}
\end{align}
This Volterra equation relates the neuronal output response $R_\phi$ to the uncondensed overlap response $R_m$. In deriving Eq.~\eqref{eq:app_volterra_RmRphi}, we used the fact that the cavity overlap is statistically equivalent to any other uncondensed overlap in the system.  A similar relation can be obtained for the uncondensed overlap autocorrelation function $C_m$ as
\begin{align}
    \begin{split}
        N &C_{m}(t,t^{\prime}) = N\langle m^{0}(t)m^{0}(t^{\prime})\rangle \\
        &=N\left\langle\int_{0}^{t}\dd\tau R_{m}(t,\tau)\eta^{0}(\tau)\int_{0}^{t^{\prime}}\dd\tau^{\prime}R_{m}(t^{\prime},\tau^{\prime})\eta^{0}(\tau^{\prime})\right\rangle  \\
        &=\int_{0}^{t}\int_{0}^{t^{\prime}}\dd\tau \dd\tau^{\prime}\,R_{m}(t,\tau)R_{m}(t^{\prime},\tau^{\prime}) C_{\phi}(\tau,\tau^{\prime})\,,
    \end{split}
\end{align}
which implies that 
\begin{align}\label{eq:app_eta0_autocorr_as_Cphi}
 \langle \eta_0(t) \eta_0(t')\rangle = \alpha \int_0^t\int_0^{t'} \dd\tau \dd\tau'R_m(t, \tau)R_m(t',\tau') C_{\phi}(\tau, \tau')\,.
\end{align}
The dynamical equation for the effective neuron cavity variable, Eq.~\eqref{eq:app_dmft_x0} constitutes, together with he Volterra equation Eq.~\eqref{eq:app_volterra_RmRphi} and the noise-noise autocorrelation function in Eq.~\eqref{eq:app_eta0_autocorr_as_Cphi}, the dynamical mean field theory of a recurrent neural network with a connectivity matrix determined by a long-term Hebbian rule, \textit{without} short-term synaptic plasticity. In the next section, we add back the short-term synaptic plasticity term and summarize the dynamical mean field theory equations.

\subsection{Final self-consistent DMFT with plastic kernel}
The plastic kernel can directly be added back to the dynamical mean field theory derived in the section above, since the neuron output autocorrelation function concentrates around its average value in the thermodynamic limit. When short-term synaptic plasticity is included back into the equation, its effects results in an additive contribution to the response term of the dynamics in Eq.~\eqref{eq:app_dmft_x0}. The final, effective dynamics of a neuron in the system is
\begin{align}\label{eq:app_dmft_x_final}
    \begin{split}
        \p_t x(t) &= -x(t) + \sum_{\mu^\ast} \xi^{\mu^\ast} m^{\mu^\ast}(t)
        \\
        &+ \int_{0}^{t} M(t,\tau)\,\phi(\tau)\,\dd\tau + \eta(t)\,,
    \end{split}
\end{align}
with a nonlinearity $\phi(t) = \tanh \gamma x(t)$. The memory kernel $M(t,\tau)$ is given by the superposition of the uncondensed overlap response function $R_m(t,\tau)$ and the short-term synaptic plasticity kernel, namely,
\begin{align}\label{eq:app_dmft_final_M}
    \begin{split}
        M(t,\tau) &\equiv \alpha R_m(t,\tau) + M_{\mathrm{syn}}(t,\tau) \\
        M_{\mathrm{syn}}(t,\tau) &\equiv  \frac{k}{p} e^{-(t-\tau)/p} C_\phi(t,\tau)\,.
    \end{split}
\end{align}
The noise $\eta(t)$ stems for the uncondensed overlaps, and it is a Gaussian noise with zero mean and autocorrelation given by
\begin{align}
\label{eq-covariance-noise-dmft}
    \begin{split}
        \langle \eta(t)\eta(t') \rangle_{\eta,\xi^{\mu^*}} &= 
    \alpha \int_{0}^{t} \! \int_{0}^{t'} 
    R_m(t,\tau)\,R_m(t',\tau') \\
    &\times C^\phi(\tau,\tau')\,\dd\tau\,\dd\tau'\,,
    \end{split}
\end{align}
where we replaced the average over the $N$-dimensional patterns $\langle\cdot\rangle$ with the average $\langle\cdot\rangle_{\eta,\xi^{\mu^*}} $ over realizations of the mean-field condensed patterns $\xi^{\mu^*}$ and the realizations of the stochastic process $\eta(t)$. The two averages are statistically equivalent in the thermodynamic limit. The uncondensed overlaps response $R_m$ is related to the neuron output response $R_\phi(t,\tau)$ through a Volterra equation, which reads
\begin{equation}\label{eq:app_dmft_Rm_final}
    R_m(t,t') = \delta(t-t') + \int_{t'}^{t} R_\phi(t,\tau)\,R_m(\tau,t')\,\dd\tau\,. 
\end{equation}
The neuron output response and autocorrelation function can be self-consistently determined from averaging over different realizations of the noise in the dynamics in Eq.~\eqref{eq:app_dmft_x_final} as
\begin{align}
    \begin{split}
        C_\phi(t,t') &= \langle \phi(t) \phi(t')\rangle_{\eta,\xi^{\mu^*}}\\
        R_\phi(t,t') &= \left\langle \frac{\delta \phi(t)}{\delta h(t')}\right\rangle_{\eta,\xi^{\mu^*}}\,.
    \end{split}
\end{align}
Finally, the \textit{condensed} overlaps can be determined self consistently as
\begin{equation}\label{eq:app_dmft_mt}
    m^{\mu^*}(t) = \langle \xi^{\mu^*} \phi(t)\rangle_{\eta,\xi^{\mu^*}}\,.
\end{equation}
Equations~\eqref{eq:app_dmft_x_final} to Eq.~\eqref{eq:app_dmft_mt} constitute the dynamical mean field theory of a recurrent neural network with Hebbian long term learning rule and short-term associative synaptic plasticity, and correspond to \cref{eq:dmft_main_x,eq:m*,eq:eta_eta,eq:Rm,eq:Rphi,eq:Cphi,eq:Kplast} of the main text, when we specialize to the case of a single condensed pattern, $s=1$ and $\xi^{1} \equiv \xi^*$.

\section{Numerical solution of the DMFT equations}
\label{app:sim_details}

In this Appendix we describe the numerical procedure used to obtain self-consistent solutions of the DMFT equations derived in Appendix~\ref{app:dmft}. The method follows the standard approach used in dynamical mean-field analysis of disordered systems~\cite{eissfeller1992new, eissfeller1994mean, roy2019numerical}: we replace the continuous-time formulation by a discretized representation on a finite time grid, express all temporal kernels as matrices acting on this grid, and iterate the resulting equations until convergence. This numerical scheme provides an explicit realization of the effective stochastic process governing a representative neuron in a network of infinitely large size, together with the associated response and correlation functions.

\subsection{Time discretization}\label{sec:app_dmft_numerics_discretization}

To simulate the DMFT equations, Eq.~\eqref{eq:app_dmft_x_final} to Eq.~\eqref{eq:app_dmft_mt}, we introduce a uniform time grid
\begin{equation}
    t_i = i\,\Delta t, 
    \qquad i = 0,\ldots,T-1\,,
\end{equation}
with time step $\Delta t$ and total horizon $T\,\Delta t$. All functions of two time variables, such as the memory kernel $M(t,\tau)$ in Eq.~\eqref{eq:app_dmft_final_M}, are then represented as $T\times T$ matrices. The entry $M_{ij}$ is then understood as an approximation to the continuous quantity $M(t_i,t_j)$. The discretization step $\Delta t$ is chosen sufficiently small that the discretized objects provide faithful approximations of their continuous-time counterparts.

The effective single-site dynamics derived in Appendix~\ref{app:dmft} is recapitulated here again for completeness:
\begin{align}
    \begin{split}
        \partial_t x(t) &= -x(t)
    + \sum_{\mu^\ast} \xi^{\mu^\ast} m^{\mu^\ast}(t)
    \\&+ \eta(t)
    + \int_0^{t} M(t,\tau)\,\phi(\tau)\,\dd\tau\,.
    \end{split}
\end{align}
To simulate this stochastic process on the discrete time grid, we employ an explicit Euler scheme. Writing $x_i := x(t_i)$\footnote{Note that with at slight abuse of notation, in this section the quantity $x_i$ refers to the membrane potential of the effective neuron described by the DMFT, evaluated at time $t_i$, rather than the membrane potential of neuron $i$ in the original network.}, the update from $t_{i-1}$ to $t_i$ takes the form
\begin{equation}
    x_i = (1 - \Delta t)\,x_{i-1} 
+ \Delta t \,\mathrm{Rate}(t_{i-1})\,,
\end{equation}
where we collect all the contribution but the linear restoring term contributions into the single term $\mathrm{Rate}(t)$, defined as
\begin{equation}
    \mathrm{Rate}(t)
    =\sum_{\mu^\ast} \xi^{\mu^\ast} m^{\mu^\ast}(t)
    + \eta(t)
    + I(t)\,,
\end{equation}
where the term $I$ is the contribution involving the memory term, namely
\begin{equation}
I(t) \equiv  \int_0^{t} M(t,\tau)\,\phi(\tau)\,\dd\tau,
\end{equation}
which represents the cumulative influence of past activity through the effective feedback kernel $M$. On the discretized time grid, this convolution is approximated by a left Riemann sum\cite{hughes2020calculus}. To simplify the notation, we introduce the matrix $M_{ij} \equiv  M(t_i,t_j)$,
which remains strictly lower-triangular because the continuous kernel $M(t,\tau)$ vanishes for $t<\tau$ by causality. The integral $I(t)$ becomes then
\begin{equation}
    I(t_{i}) \approx 
\sum_{j = 0}^{i} M_{ij}\,\phi_j\,\Delta t.
\end{equation}
The discretization converts the continuous-time DMFT equations into a set of coupled algebraic updates that can be iterated forward in time. All remaining components of the algorithm—the evaluation of response functions, noise covariance, and self-consistent kernels—are built upon this same discretized temporal structure.

\subsection{Single-site response and $R_\phi$}

To complete the DMFT description of the effective neuron, we must evaluate the linear response of the process $x(t)$ to an infinitesimal perturbation applied at an earlier time $\tau$. This quantity is denoted by
\begin{equation}
    R_x(t,\tau) \equiv \left\langle\frac{\delta x(t)}{\delta h(\tau)}\right\rangle_{\eta,\xi^*}\,,
\end{equation}
and it measures the sensitivity of the state at time $t$ to a small auxiliary field $h(\tau)$ coupled to the neuron at time $\tau$. As in standard dynamical mean-field theory, the response of the output variable $\phi(t)$ then follows by multiplying $R_x$ by the instantaneous gain $\phi'(x)$.
It is convenient to introduce first the \emph{single-trajectory} susceptibility
\begin{equation}
R_x^{(r)}(t,\tau)
\equiv \frac{\delta x^{(r)}(t)}{\delta h(\tau)} ,
\end{equation}
for a given realization $a$ of the effective stochastic process (noise and condensed pattern). The disorder-averaged response is then obtained as
\begin{equation}
    R_x(t,\tau) = \frac{1}{M}\sum_{r=1}^M R_x^{(r)}(t,\tau) .
\end{equation}
From Eq.~\eqref{eq:app_dmft_x_final} we see that $R^{(r)}_x$ satisfies the following linear Volterra equation,
\begin{align}
\begin{split}
\partial_t R^{(r)}_x(t,\tau)
&= - R^{(r)}_x(t,\tau) + \delta(t-\tau)\\
&+ \int_{\tau}^{t}
M(t,\tau') \phi'\left(x^{(r)}(\tau')\right)R^{(r)}_x(\tau',\tau)\,\dd\tau' ,
\label{eq:app_Sx_volterra}
\end{split}
\end{align}

with the causal condition $\lim_{\tau\to t_-} R^{(r)}_x(t,\tau)=1$. For the numerical solution we use the same uniform time grid $t_i = i\Delta t$. Approximating the time derivative in Eq.~\eqref{eq:app_Sx_volterra} by a forward Euler step and the memory integral by a left Riemann sum we obtain, for $t_i>t_j$,
\begin{align}
\begin{split}
R_x^{(r)}&(t_i,t_j)
\approx (1-\Delta t)\,R_x^{(r)}(t_{i-1},t_j)\\
&+ \Delta t^2 \sum_{k=j}^{i-1}
M_{i-1,k}\,\phi'\bigl(x^{(r)}(t_k)\bigr)\,
R_x^{(r)}(t_k,t_j)\\
&+ \delta_{i-1,j},
\label{eq:app_Sx_discrete_scalar}
\end{split}
\end{align}
with $R_x^{(r)}(t_i,t_i)=0$. Here $M_{i-1,k} \equiv M(t_{i-1},t_k)$ denotes the discrete memory kernel, and the Dirac delta has been replaced by a Kronecker delta,
$\Delta t\,\delta(t_{i-1}-t_j) = \delta_{i-1,j}$.

In the implementation, we evolve Eq.~\eqref{eq:app_Sx_discrete_scalar} separately for each source time $t_j$. Given the single-trajectory susceptibilities $R_x^{(r)}(t_i,t_s)$ and the corresponding gains $\phi'^{(r)}_i$, we compute the output response by averaging over $M$ independent trajectories:
\begin{equation}
    R_\phi(t_i,t_s)
    \approx \frac{1}{M}\sum_{r=1}^M
    \phi'^{(r)}_i\,R_x^{(r)}(t_i,t_s).
\end{equation}
In matrix form, writing
\begin{equation}
    (\mathbf{R}_\phi)_{i,j} \equiv R_\phi(t_i,t_j),
\end{equation}
the numerical estimate used in the simulations is
\begin{equation}
(\mathbf{R}_\phi)_{i,s}
\approx \frac{1}{M}\sum_{r=1}^M
\phi'^{(r)}_i\,R_{x,i}^{(r)}.
\label{eq:app_Rphi_mc}
\end{equation}
The resulting matrix $\mathbf{R}_\phi$ is strictly lower-triangular (up to numerical noise), consistent with causality, and in practice we enforce this structure by projecting out any spurious entries above the diagonal. The matrix $\mathbf{R}_\phi$ is then used as the continuous $O(1)$ kernel in the construction of the Hopfield response $R_m$ and the corresponding noise covariance, as discussed in Appendix~\ref{app:discretization_Rm}.

\subsection{Discretization of $R_m$ and the Hopfield kernel}\label{app:discretization_Rm}

The response of the uncondensed overlaps, denoted $R_m(t,t')$, plays a central role in the DMFT equations. As shown in Eq.~\eqref{eq:app_dmft_Rm_final}, it satisfies a linear Volterra equation involving the output response $R_\phi$, which we recapitulate here:
\begin{equation}
    R_m(t,t') = \delta(t-t') 
    + \int_{t'}^{t} R_\phi(t,\tau)\,R_m(\tau,t')\,\dd\tau\,.
\end{equation}
To evaluate $R_m$ numerically, we discretize the integral equation on the same time grid used throughout the simulation. Replacing the time integral by a left Riemann sum yields
\begin{align}
R_m(t_i,t_j)
&\approx \frac{\delta_{i,j}}{\Delta t}
+ \sum_{k=j}^{i-1} R_\phi(t_i,t_k)\,R_m(t_k,t_j)\,\Delta t\,,
\label{eq:Rm-disc-kernel}
\end{align}
where the factor $\Delta t^{-1}$ appears because the Dirac delta is replaced by a Kronecker delta on the grid. This expression is still a Volterra-type recursion: the value of $R_m$ at time $t_i$ depends only on earlier times, ensuring strict causality. It is convenient to introduce an operator-valued representation that mirrors the continuous-time treatment. To this end, we define $
(\mathbf{R}_m^{\mathrm{op}})_{i,s}
:= R_m(t_i,t_s)\,\Delta t$,
so that $\mathbf{R}_m^{\mathrm{op}}$ is a strictly lower-triangular matrix with entries of order $O(1)$ as $\Delta t\to 0$. Multiplying Eq.~\eqref{eq:Rm-disc-kernel} by $\Delta t$ and defining analogously $(\mathbf{R}_\phi^{\mathrm{op}})_{i,j}
:= R_\phi(t_i,t_j)\,\Delta t,$, we obtain
\begin{align}
(\mathbf{R}_m^{\mathrm{op}})_{i,j}
&= \delta_{i,j}
+ \sum_{k=j}^{i-1}
(\mathbf{R}_\phi^{\mathrm{op}})_{i,k}\,
(\mathbf{R}_m^{\mathrm{op}})_{k,j}.
\end{align}
In matrix notation, this equation has the following more compact form,
\begin{align}\label{eq:app_Rm_as_R_phi}
\mathbf{R}_m^{\mathrm{op}}
= \bigl(\mathbf{I}_T - \mathbf{R}_\phi^{\mathrm{op}}\bigr)^{-1}\,,
\end{align}
where $\mathbf{I}_T$ is a $T\times T$ identity matrix. The original kernel $\mathbf{R}_m$ (without the $\Delta t$ normalization) can be then reconstructed through
\begin{align}\label{eq:app_Rm_as_Ropm}
\mathbf{R}_m = \frac{1}{\Delta t}\,\mathbf{R}_m^{\mathrm{op}}\,.
\end{align}
The matrix $\mathbf{R}^\text{op}_\phi$ can be computed from Eq.~\eqref{eq:app_Rphi_mc}. \cref{eq:app_Rphi_mc,eq:app_Rm_as_R_phi,eq:app_Rm_as_Ropm} constitute the core of the response computation loop in the numerical solution of the DMFT. 

\subsection{Noise covariance}
 In Appendix~\ref{app:dmft} we derived the continuous-time expression for the noise covariance, recapitulated here as
\begin{equation}\label{eq:app_Ceta_repeated}
    C_\eta(t,t')
    = \alpha \int_0^{t}\!\!\int_0^{t'} 
    R_m(t,\tau)\,C_\phi(\tau,\tau')\,R_m(t',\tau')
    \,\dd\tau\,\dd\tau'
\end{equation}
On the discretized time grid, the double integral is replaced by a double Riemann sum. Using the discretized notation, Eq.~\eqref{eq:app_Ceta_repeated} becomes
\begin{align}
C_\eta(t_i,t_j)
&\approx \alpha
\sum_{k,l}
R_m(t_i,t_k)\,C_\phi(t_k,t_l)\,R_m(t_j,t_l)\Delta t^2.
\end{align}
This expression is the discrete analogue of the continuous convolution structure: the noise at two times $t_i$ and $t_k$ is correlated whenever the uncondensed modes can jointly propagate fluctuations from earlier times $t_i$ and $t_j$. We rewrite this expression using the normalized operators normalization introduced in Sec.~\ref{app:discretization_Rm}, $(\mathbf{R}_m^{\mathrm{op}})_{i,j}
= R_m(t_i,t_j)\,\Delta t$, and obtain
\begin{align}
    \begin{split}
        C_\eta(t_i,t_k)
        &\approx
        \alpha \sum_{j,l}
        (\mathbf{R}_m^{\mathrm{op}})_{i,j}\,
        C_\phi(t_j,t_l)\,
        (\mathbf{R}_m^{\mathrm{op}})_{k,l}.
    \end{split}
\end{align}
The powers of $\Delta t$ cancel out, leaving a finite expression that remains well-behaved as $\Delta t \to 0$. In matrix notation, the expression above takes the compact form
\begin{align}
\mathbf{C}_\eta
\approx
\alpha\,\mathbf{R}_m^{\mathrm{op}}\,\mathbf{C}_\phi\,(\mathbf{R}_m^{\mathrm{op}})^\top.
\end{align}
The resulting matrix is guaranteed to be symmetric and positive semi-definite—properties required of a covariance matrix. The matrix $\mathbf{C}_\eta$ is then used to generate the Gaussian noise trajectories needed to simulate the effective process. Specifically, for each DMFT iteration we draw independent realizations of $\eta(t)$ with covariance $\mathbf{C}_\eta$, ensuring that the statistics of the noise affecting the effective neuron are consistent with those of the full network in the thermodynamic limit.

\subsection{Main algorithm}

We now summarize the numerical procedure used to obtain self-consistent solutions of the DMFT equations. Each iteration consists of three stages: (i) constructing the operators and memory kernels from the current correlation and response estimates; (ii) sampling and integrating the effective single-neuron dynamics; and (iii) updating the different dynamical order parameters. The process is repeated until convergence. We focus, without loss of generality, to the case of a single condensed overlap

\begin{enumerate}
    \item \textbf{Initialization.}
    We begin from simple initial guesses for the autocorrelation, response, and condensed overlap:
    \begin{align}
        \begin{split}
            \mathbf{C}_\phi &\leftarrow \mathbf{I}_T\,,\\
            \mathbf{R}_\phi &\leftarrow \mathbf{0}\,, \\
            m(t_i) &\leftarrow 0\,.
        \end{split}
    \end{align}
    \item \textbf{Construction of operators and kernels.}
    Using the discretized operators defined in the preceding subsections, we obtain the following set of operators:
    \begin{align}\label{eq:app_update_CmR}
        \begin{split}
            \mathbf{R}_\phi^{\mathrm{op}} &\leftarrow \mathbf{R}_\phi\,\Delta t\,, \\
            \mathbf{R}_m^{\mathrm{op}} &\leftarrow (\mathbf{I}_T - \mathbf{R}_\phi^{\mathrm{op}})^{-1}\,,\\
            \mathbf{C}_m &\leftarrow \alpha\,\mathbf{R}_m^{\mathrm{op}}\,\mathbf{C}_\phi\,(\mathbf{R}_m^{\mathrm{op}})^\top\,.
        \end{split}
    \end{align}
    The plasticity contribution is constructed directly from the current output autocorrelation:
    \begin{align}
        (K_{\mathrm{plast}})_{ij}
        =
        \begin{cases}
            \dfrac{k}{p}\,e^{-(t_i - t_j)/p}\,C_\phi(t_i,t_j), & i>j,\\[4pt]
            0, & i\le j,
        \end{cases}
    \end{align}
    and the total memory kernel used in the dynamics is
    \begin{align}
        \mathbf{W} \leftarrow \alpha \mathbf{R}_m + \mathbf{K}_{\mathrm{plast}}\,.
    \end{align}

    \item \textbf{Sampling of trajectories.}
    For each iteration we generate $M$ independent effective-neuron trajectories:
    \begin{enumerate}
        \item Draw Gaussian noise samples with the computed covariance,
        \begin{align}
            \beeta^{(r)} \sim \mathcal{N}(0,\mathbf{C}_\eta),\qquad r=1,\dots,M\,, 
        \end{align}
        where $\beeta^{(r)}$ is a $T$-dimensional vector containing one realization of a noise trajectory.
        \item Sample i.i.d.\ condensed-pattern components $\xi^{(r)}\in\{\pm1\}$ from a uniform distribution.
        \item Initialize each trajectory around the condensed direction.  
        For an initial amplitude $g$ and desired initial overlap $a_{\mathrm{bar}}$, set
        \begin{align}\label{eq:app_a}
            \begin{split}
                a &= g\,a_{\mathrm{bar}}\,,\\
                \sigma_z^2 &= g^2 - a^2\,,
            \end{split}
        \end{align}
        and draw
        \begin{align}\label{eq:app_x0_init}
            \begin{split}
                x_0^{(r)} &= a\,\xi^{(r)} + z_0^{(r)}\,,\\
                \phi_0^{(r)} &= \phi(x_0^{(r)})\,,
            \end{split}
        \end{align}
        where $z_0^{(r)} \sim \mathcal{N}(0,\sigma_z^2)$ is a set of independent, identically distributed Gaussian random numbers with zero mean and variance $\sigma^2_z$. 
    \end{enumerate}

    \item \textbf{Forward integration.}
    For each realization $r$ and each time index $i=1,\dots,T-1$, we integrate the discretized dynamics using the Euler scheme discussed in Sec.~\ref{sec:app_dmft_numerics_discretization}, namely
    \begin{align}
        \begin{split}
            x_i^{(r)} 
            &= (1-\Delta t)\,x_{i-1}^{(r)}
            + \Delta t\left[\eta_{i-1}^{(r)} + \xi^{r}m_{i-1}\right] \\
            &+\Delta t^2 \sum_{j=0}^{i-1} W_{i,j}\phi^{(r)}_j\,,
        \end{split}
    \end{align}
    with $\phi^{(r)}_j \equiv \phi\left(x_j^{(r)}\right)$
    where $\mathrm{Rate}_{\mathbf{W}}$ incorporates the retarded interactions through the memory kernel $\mathbf{W}$ and the sampled noise $\eta^{(r)}$.  
    We then update $\phi_i^{(r)}$.

    \item \textbf{Estimation of overlap and autocorrelations}
    The new values of the neuron output autocorrelation function, $C_\phi^{\text{new}}(t_i,t_j)$, and of the condensed overlap, $m^\text{new}(t_i)$, are estimated from the ensemble of trajectories:
    \begin{align}
        C_\phi^{\mathrm{new}}(t_i,t_j)
        &= \frac{1}{M}\sum_{r=1}^M \phi_i^{(r)}\,\phi_j^{(r)}, \nonumber\\
        m^{\mathrm{new}}(t_i)
        &= \frac{1}{M}\sum_{r=1}^M \xi^{(r)}\,\phi_i^{(r)}. \nonumber
    \end{align}

    \item \textbf{Estimation of the response.}
    For each source time $t_j$, we propagate the response function $R_x^{(r)}(\cdot,j)$ along every trajectory using the discretized response equation given by Eq.~\eqref{eq:app_Sx_discrete_scalar}.  
    The output response is then
    \begin{align}
        R_\phi^{\mathrm{new}}(t_i,t_j)
        = \frac{1}{M}\sum_{r=1}^M
        \phi'\left(x_i^{(r)}\right)\,R_x^{(r)}(t_i,t_j). 
    \end{align}

    \item \textbf{Damped update.}
    To stabilize the fixed-point iteration, we update the order parameters using a linear interpolation between the old order parameters and the new order parameters, using an injection parameter $\gamma_I\in(0,1)$:
    \begin{align}
        \begin{split}
            \mathbf{C}_\phi &\leftarrow (1-\gamma_I)\mathbf{C}_\phi + \gamma_I\,\mathbf{C}_\phi^{\mathrm{new}}\,,\\
            m(t_i) &\leftarrow (1-\gamma_I)m(t_i) + \gamma_I\,m^{\mathrm{new}}(t_i)\,,\\
            \mathbf{R}_\phi &\leftarrow (1-\gamma_I)\mathbf{R}_\phi + \gamma_I\,\mathbf{R}_\phi^{\mathrm{new}}\,,
        \end{split}
    \end{align}
    the responses and correlations of of uncondensed overlap, $\mathbf{R}_m$, and $\mathbf{C}_m$, are then computed using Eq.~\eqref{eq:app_update_CmR}.
    
    \item \textbf{Convergence.}
    Steps $3$-$7$ are repeated until the relative changes in $\mathbf{C}_\phi$, $\mathbf{R}_\phi$, and $m(t)$ fall below a chosen tolerance.  Additionally, we track the normalized overlap
    \begin{align}
        \overline m(t_i) = \frac{m(t_i)}{\sqrt{C_\phi(t_i,t_i)}} \nonumber
    \end{align}
    over the final portion of the time window to verify that the retrieval dynamics has entered a stable (or asymptotic) regime.
\end{enumerate}

The results shown in the paper are obtained using $\Delta t=0.25$, $\gamma_I=0.02$, $\gamma \approx3.4$ and 2048 independent trajectories. We simulated for 8000 iterations, after which the maximum relative displacement of the order parameters was below $10^{-3}$.
\section{Finite size simulations}\label{app:finite_size_sim}

\begin{figure}
    \centering
\includegraphics[width=\linewidth]{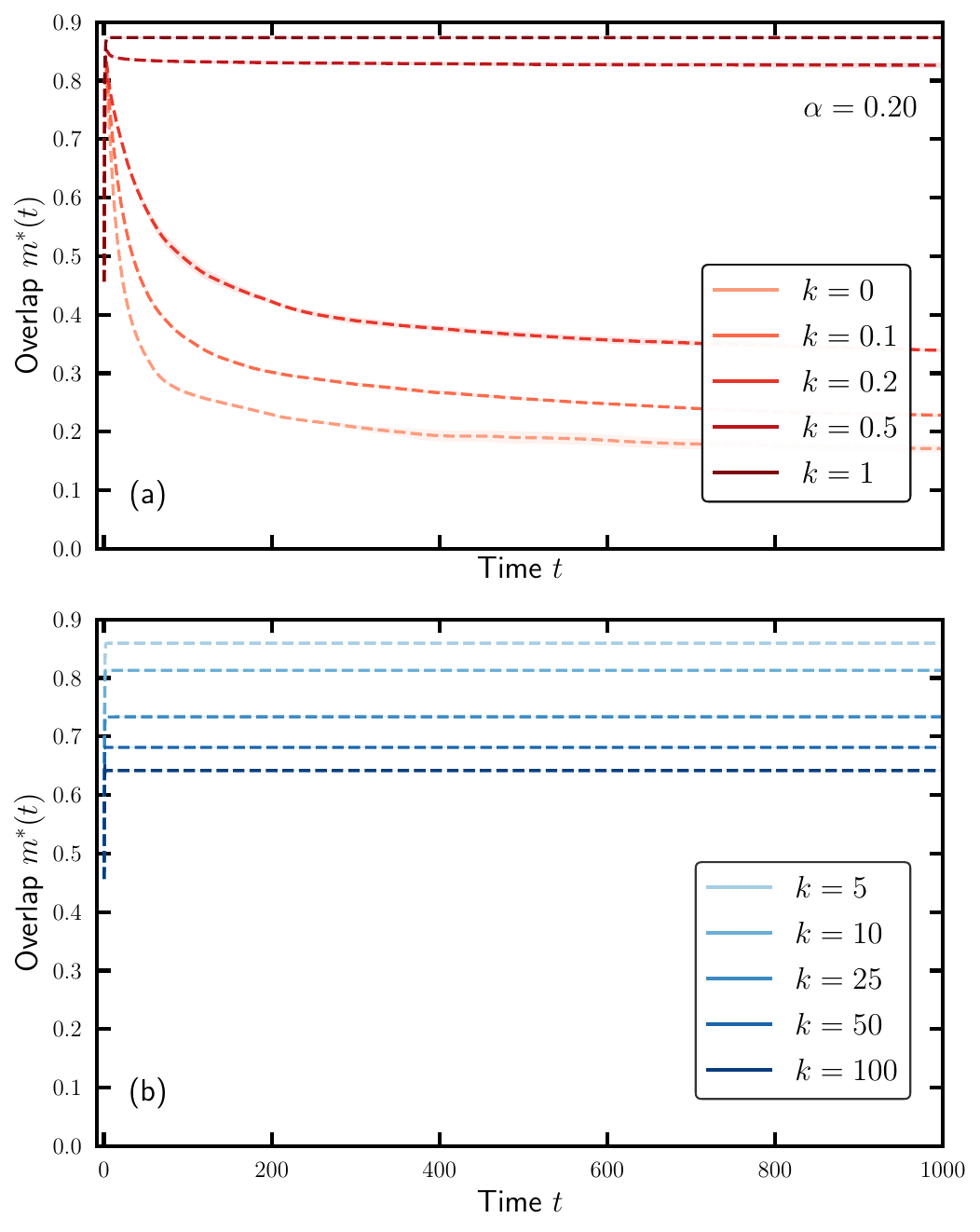}
    \caption{\textbf{Finite size simulations of the long-time dynamics of the neural network.} We show the temporal evolution of the overlap $m^*(t) $against a target pattern for a network of $N=20000$ neurons for different values of the short-term synaptic plasticity strength $k$.}
    \label{fig:long_time_finite_size}
\end{figure}

The finite size simulations presented in Fig.~\ref{fig:dmft_vs_sims} are obtained by solving numerically \cref{eq:neuron_dyn,eq:stp_dyn}, for a system of $N=20000$ neurons, using Euler algorithm with a time-step $\Delta t=0.25$. The dynamical overlap is computed from the average $m^*(t) = \frac{1}{N}\sum_{i=1}^N \langle \xi^*_i \phi(x_i(t))\rangle$, where the average $\langle \cdot \rangle$ is obtained by averaging over $2048$ replicates with different realizations of the target overlap. The target pattern is conventionally chosen to be the first one generated when constructing the matrix $J_{ij}$, $\xi^*_i \equiv \xi^1_i$. The network is initialized using the same procedure described in \cref{eq:app_a,eq:app_x0_init}. We show in Fig. \ref{fig:long_time_finite_size} the temporal evolution of the overlap $m^*(t)$ at large times for a finite-size system, both with and without short-term synaptic plasticity. 
The overlap $m^*(t)$ is measured as
\begin{equation}
    m^*(t) = \frac 1 N \sum_{i=1}^N \langle \xi^*_i \phi_i(t)\rangle\,, 
\end{equation}
where the brackets $\langle\cdot\rangle$ denote an average over different realizations of the Hopfield matrix $J_{ij}$, and different realizations of the initial condition for the population of neurons.

\section{Jacobian stability analysis}
\label{app:jacobian}

To characterize the stability of a stationary state of the joint neural--synaptic dynamics, we examine the linear response of the full system around a fixed point. The model evolves in the combined space of neuronal activations and plastic synaptic weights, a space of dimension $(N + N^2)$. A generic state is specified by the vector of post-synaptic potentials $\mathbf{x} \in \mathbb{R}^N$ together with the plastic weight matrix $\mathbf{A} \in \mathbb{R}^{N\times N}$. Their evolution equations read
\begin{align}\label{eq:app_dynamics_appE}
    \begin{split}
        \dot{x}_i &= -x_i + \sum_{j=1}^N (J_{ij} + A_{ij})\,\phi_j\,,\\
        \dot{A}_{ij} &= \frac{1}{p}\!\left(-A_{ij} + \frac{k}{N}\,\phi_i\phi_j\right)\,,
    \end{split}
\end{align}
where $\phi_i=\phi(x_i)$ is the neuronal transfer function. At a fixed point $(x^*,A^*)$, both neuronal activity and synapses are stationary, and the plastic component of the weights takes the Hebbian-like form
\begin{equation}
A_{ij}^* = \frac{k}{N}\phi_i^*\phi_j^*\,.
\end{equation}
For later convenience we introduce the diagonal matrix $\Phi'$ with entries $\Phi'_{ij} = \phi'_i \delta_{ij}$, the derivative of the activation function evaluated at $x_i^*$. Thus, unless stated otherwise, every occurrence of $\phi'_i$ or $\Phi'$ refers to derivatives evaluated at $x_i = x_i^*$.

The Jacobian of the dynamics of the full system, $\mathcal{J}$, at the fixed point can be written as a block matrix coupling neural and synaptic degrees of freedom:
\begin{align}
\mathcal{J}
\equiv\begin{pmatrix}
\frac{\p \dot x_i}{\p x_j} & \frac{\p \dot x_{i}}{\p A_{jk}} \\
\frac{\p \dot A_{ij}}{\p x_k} & \frac{\p \dot A_{ij}}{\p A_{kl}}
\end{pmatrix}\equiv 
\begin{pmatrix}
\mathcal{J}_{xx} & \mathcal{J}_{xA} \\
\mathcal{J}_{Ax} & \mathcal{J}_{AA}
\end{pmatrix}\,,
\end{align}
and each block may be computed by direct differentiation of the dynamics in Eq.~\eqref{eq:app_dynamics_appE}. They read respectively,
\begin{align}
    \begin{split}
        [\mathcal{J}_{xx}]_{ij} &= -\delta_{ij} + \sum_{k=1}^N W^*_{ik} \Phi'_{kj}\,,\\
        [\mathcal{J}_{xA}]_{i,jk} &= \delta_{ij}\phi_k\,,\\
        [\mathcal{J}_{Ax}]_{ij,k} &= \frac{k}{pN}\!\left(\phi'_i \phi_j\delta_{ik}
        + \phi_i\phi'_j \delta_{jk}\right)\,, \\
        [\mathcal{J}_{AA}]_{ij,kl} &= -\frac{1}{p}\delta_{ik}\delta_{jl}\,.
    \end{split}
\end{align}

The eigenvalues of the Jacobian of the dynamics of the composite neurons-synapses system solve the characteristic equation
\begin{equation}
    \det(\mathcal{J}-\lambda \mathbf{I})=0.
\end{equation}
A simplification arises from the structure of $\mathcal{J}_{AA}$. Indeed, let us observe that
\begin{equation}
    \mathcal{J}_{AA}-\lambda \mathbf{I} = -(\lambda + 1/p) \mathbf{I}_{N^2}\,,
\end{equation}
where $\mathbf{I}_{N^2}$ is an $N^2\times N^2$ identity matrix. The matrix  $\mathcal{J}_{AA}-\lambda \mathbf{I}$ is thus invertible whenever $\lambda \neq -1/p$. This determines a set of degenerate eigenvalues, equal to $-1/p$. For the other  eigenvalues, we can use the determinant identity for block matrices:
\begin{align}\label{eq:app_det_block}
\det(\mathcal{J}-\lambda \mathbf{I})
= \det(\mathcal{J}_{AA}-\lambda \mathbf{I})\;
  \det\!\left( \mathbf{S}(\lambda) \right)\,, 
\end{align}
where the stability-relevant part of the dynamics is captured by the $N\times N$ Schur complement $S(\lambda)$, defined as
\begin{align}
    \begin{split}
        \mathbf{S}(\lambda)
        &\equiv (\mathcal{J}_{xx}-\lambda \mathbf{I})
        - \mathcal{J}_{xA}(\mathcal{J}_{AA}-\lambda \mathbf{I})^{-1}\mathcal{J}_{Ax} \nonumber\\
        &= (\mathcal{J}_{xx}-\lambda \mathbf{I})
        + \frac{1}{\lambda+1/p} \mathbf{C}\,,
    \end{split}
\end{align}
where the matrix $\mathbf{C}$ is defined as
\begin{equation}
    \mathbf{C} \equiv \mathcal{J}_{xA}\,\mathcal{J}_{Ax}\,,
\end{equation}
and it is a $N\times N$ matrix that encodes the linear feedback from synaptic to neural perturbations. Inserting the explicit expressions gives
\begin{align}
    \begin{split}
        C_{ij}
        &= \sum_{k,l} 
        (\mathcal{J}_{xA})_{i,(k,l)}\,
        (\mathcal{J}_{Ax})_{(k,l),j}\\
        &= \frac{k \phi'_j}{pN}
        \left[
        \delta_{ij}\!\left(\sum_m \phi_m^2\right)
        + \phi_i\phi_j
        \right]\,. 
    \end{split}
\end{align}
Introducing the scalar
\begin{equation}
    N q = \sum_m (\phi_m^*)^2,
\end{equation}
we obtain the compact expression
\begin{align}
\mathbf{C} = \frac{k}{p}\,\Phi'\,(q \mathbf{I} + \frac{1}{N}\bphi^*\otimes \bphi^*)\,. 
\end{align}

From Eq.~\eqref{eq:app_det_block}, we see that the nontrivial eigenvalues of $\mathcal{J}$ satisfy the reduced condition
\begin{align}
\det\!\left(
(\mathcal{J}_{xx}-\lambda \mathbf{I})
+ \frac{1}{\lambda+1/p}\mathbf{C}
\right)=0\,. 
\end{align}
Multiplying both sides of this equation by $(\lambda+1/p)^N$ yields the equivalent quadratic eigenvalue problem
\begin{align}
\label{structure-equation}
\det\!\bigl((\lambda+1/p)(\mathcal{J}_{xx}-\lambda \mathbf{I})+\mathbf{C}\bigr)=0\,.
\end{align}
This equation produces the $2N$ eigenvalues that couple neural and synaptic fluctuations.  
The remaining $N^2-N$ eigenvalues are the trivial ones, $\lambda = -\frac{1}{p}$,
stemming directly from the diagonal block $\mathcal{J}_{AA}$. The remaining $N^2-N$ eigenvalues come directly from the synaptic block
$\mathcal{J}_{AA} = -(1/p)\,\mathbf{I}_{N^2}$. For an eigenvalues $\lambda = -1/p$, the eigenvectors belong entirely to the space of synaptic connectivity, and do not involve the neuron degrees of freedom. In this case, the perturbation to the plastic synapses must be a null eigenvector of $\mathcal{J}_{xA}$. Since $\mathcal{J}_{xA}$ has rank $N$,
its kernel has dimension $N^2-N$, providing $N^2-N$ linearly independent
eigenvectors with eigenvalue $\lambda=-1/p$.

For numerical implementations, we proceed as follows. The reduced eigenvalue condition in Eq.~\eqref{structure-equation} can be rewritten as a quadratic matrix polynomial in $\lambda$. Expanding the term inside the determinant gives
\begin{align}
&(\lambda+1/p)\mathcal{J}_{xx} - (\lambda+1/p)\lambda \mathbf{I} + \mathbf{C} \nonumber\\
&= -\lambda^2 \mathbf{I}
+ \lambda\!\left(\mathcal{J}_{xx}-\frac{1}{p}\mathbf{I}\right)
+ \left(\frac{1}{p}\mathcal{J}_{xx}+\mathbf{C}\right).
\end{align}
Multiplying by $-1$ (which does not change the roots of the determinant) we obtain the equivalent quadratic eigenvalue problem
\begin{equation}
\det\!\bigl(\lambda^2 \mathbf{I} + \lambda \mathbf{A} + \mathbf{B}\bigr)=0\,,
\end{equation}
with
\begin{equation}
\mathbf{A} = \frac{1}{p}\mathbf{I} - \mathcal{J}_{xx},
\qquad
\mathbf{B} = -\frac{1}{p}\mathcal{J}_{xx} - \mathbf{C}.
\end{equation}

To solve this quadratic problem using standard linear-algebra routines, we embed it into a linear eigenvalue problem on a doubled space via the $2N\times 2N$ companion matrix
\begin{equation}
\mathbf{M}_{\mathrm{comp}}
=
\begin{pmatrix}
\mathbf{0} & \mathbf{I}_N \\
-\mathbf{B} & -\mathbf{A}
\end{pmatrix}.
\end{equation}
We then want to solve the problem
\begin{align}
    \mathbf{M}_{\text{comp}} \mathbf{v} = \lambda \mathbf{v}
\end{align}
It is possible to show that $\lambda$ is a root of Eq.~\eqref{structure-equation} if and only if $\lambda$ is an eigenvalue of $\mathbf{M}_{\text{comp}}$.

In summary, the $2N$ eigenvalues of the companion matrix $\mathbf{M}_{\mathrm{comp}}$ coincide exactly with the $2N$ nontrivial eigenvalues obtained from Eq.~\eqref{structure-equation}, i.e.\ the coupled neuron--synapse modes of the full Jacobian. In the numerical implementation we therefore construct $\mathbf{M}_{\mathrm{comp}}$ and diagonalize it with a standard eigenvalue solver, while the remaining $N^2-N$ eigenvalues are fixed at $\lambda=-1/p$ from the synaptic block $\mathcal{J}_{AA}$.

\bibliographystyle{apsrev4-2}
\bibliography{biblio}

@article{kuhn1993,
  title={Statistical mechanics for neural networks with continuous-time dynamics},
  author={K{\"u}hn, R and B{\"o}s, S},
  journal={Journal of Physics A: Mathematical and General},
  volume={26},
  number={4},
  pages={831},
  year={1993},
  publisher={IOP Publishing}
}

@article{sompolinsky1981dynamic,
  title={Dynamic theory of the spin-glass phase},
  author={Sompolinsky, Haim and Zippelius, Annette},
  journal={Physical Review Letters},
  volume={47},
  number={5},
  pages={359},
  year={1981},
  publisher={APS}
}

@article{sompolinsky1982relaxational,
  title={Relaxational dynamics of the Edwards-Anderson model and the mean-field theory of spin-glasses},
  author={Sompolinsky, Haim and Zippelius, Annette},
  journal={Physical Review B},
  volume={25},
  number={11},
  pages={6860},
  year={1982},
  publisher={APS}
}

@article{tokita1994replica,
  title={The replica-symmetry-breaking solution of the Hopfield model at zero temperature: critical storage capacity and frozen field distribution},
  author={Tokita, K},
  journal={Journal of Physics A: Mathematical and General},
  volume={27},
  number={13},
  pages={4413},
  year={1994},
  publisher={IOP Publishing}
}

@article{du2024active,
  title={Active oscillatory associative memory},
  author={Du, Matthew and Behera, Agnish Kumar and Vaikuntanathan, Suriyanarayanan},
  journal={The Journal of Chemical Physics},
  volume={160},
  number={5},
  year={2024},
  publisher={AIP Publishing}
}

@article{zucker2002short,
  title={Short-term synaptic plasticity},
  author={Zucker, Robert S and Regehr, Wade G},
  journal={Annual review of physiology},
  volume={64},
  number={1},
  pages={355--405},
  year={2002},
  publisher={Annual Reviews 4139 El Camino Way, PO Box 10139, Palo Alto, CA 94303-0139, USA}
}

@article{volianskis2013different,
  title={Different NMDA receptor subtypes mediate induction of long-term potentiation and two forms of short-term potentiation at CA1 synapses in rat hippocampus in vitro},
  author={Volianskis, Arturas and Bannister, Neil and Collett, Valerie J and Irvine, Mark W and Monaghan, Daniel T and Fitzjohn, Stephen M and Jensen, Morten S and Jane, David E and Collingridge, Graham L},
  journal={The Journal of physiology},
  volume={591},
  number={4},
  pages={955--972},
  year={2013},
  publisher={Wiley Online Library}
}

@article{malenka1991postsynaptic,
  title={Postsynaptic factors control the duration of synaptic enhancement in area CA1 of the hippocampus},
  author={Malenka, Robert C},
  journal={Neuron},
  volume={6},
  number={1},
  pages={53--60},
  year={1991},
  publisher={Cell Press}
}

@article{roy2019numerical,
  title={Numerical implementation of dynamical mean field theory for disordered systems: Application to the Lotka--Volterra model of ecosystems},
  author={Roy, Felix and Biroli, Giulio and Bunin, Guy and Cammarota, Chiara},
  journal={Journal of Physics A: Mathematical and Theoretical},
  volume={52},
  number={48},
  pages={484001},
  year={2019},
  publisher={IOP Publishing}
}

@article{gao2015simplicity,
  title={On simplicity and complexity in the brave new world of large-scale neuroscience},
  author={Gao, Peiran and Ganguli, Surya},
  journal={Current opinion in neurobiology},
  volume={32},
  pages={148--155},
  year={2015},
  publisher={Elsevier}
}

@book{hughes2020calculus,
  title={Calculus: Single and multivariable},
  author={Hughes-Hallett, Deborah and Gleason, Andrew M and McCallum, William G},
  year={2020},
  publisher={John Wiley \& Sons}
}

@article{eissfeller1994mean,
  title={Mean-field Monte Carlo approach to the Sherrington-Kirkpatrick model with asymmetric couplings},
  author={Eissfeller, H and Opper, M},
  journal={Physical Review E},
  volume={50},
  number={2},
  pages={709},
  year={1994},
  publisher={APS}
}

@article{eissfeller1992new,
  title={New method for studying the dynamics of disordered spin systems without finite-size effects},
  author={Eissfeller, H and Opper, M},
  journal={Physical review letters},
  volume={68},
  number={13},
  pages={2094},
  year={1992},
  publisher={APS}
}

@article{amit1987statistical,
  title={Statistical mechanics of neural networks near saturation},
  author={Amit, Daniel J and Gutfreund, Hanoch and Sompolinsky, Haim},
  journal={Annals of physics},
  volume={173},
  number={1},
  pages={30--67},
  year={1987},
  publisher={Elsevier}
}

@article{kuhn1991statistical,
  title={Statistical mechanics for networks of graded-response neurons},
  author={K{\"u}hn, Reimer and B{\"o}s, Siegfried and van Hemmen, Jan L},
  journal={Physical Review A},
  volume={43},
  number={4},
  pages={2084},
  year={1991},
  publisher={APS}
}

@article{del2014cavity,
  title={Cavity method: Message passing from a physics perspective},
  author={Del Ferraro, Gino and Wang, Chuang and Mart{\'\i}, Dani and M{\'e}zard, Marc},
  journal={arXiv preprint arXiv:1409.3048},
  year={2014}
}

@inproceedings{hinton1987using,
  title={Using fast weights to deblur old memories},
  author={Hinton, Geoffrey E and Plaut, David C},
  booktitle={Proceedings of the ninth annual conference of the Cognitive Science Society},
  pages={177--186},
  year={1987}
}

@article{gardner1987multiconnected,
  title={Multiconnected neural network models},
  author={Gardner, Elizabeth},
  journal={Journal of Physics A: Mathematical and General},
  volume={20},
  number={11},
  pages={3453},
  year={1987},
  publisher={IOP Publishing}
}

@article{lucibello2024exponential,
  title={Exponential capacity of dense associative memories},
  author={Lucibello, Carlo and M{\'e}zard, Marc},
  journal={Physical Review Letters},
  volume={132},
  number={7},
  pages={077301},
  year={2024},
  publisher={APS}
}

@article{horn1988capacities,
  title={Capacities of multiconnected memory models},
  author={Horn, David and Usher, Marius},
  journal={Journal de Physique},
  volume={49},
  number={3},
  pages={389--395},
  year={1988},
  publisher={Soci{\'e}t{\'e} Fran{\c{c}}aise de Physique}
}

@article{ros2019complex,
  title={Complex energy landscapes in spiked-tensor and simple glassy models: Ruggedness, arrangements of local minima, and phase transitions},
  author={Ros, Valentina and Ben Arous, Gerard and Biroli, Giulio and Cammarota, Chiara},
  journal={Physical Review X},
  volume={9},
  number={1},
  pages={011003},
  year={2019},
  publisher={APS}
}

@article{abbott1987storage,
  title={Storage capacity of generalized networks},
  author={Abbott, Laurence F and Arian, Yair},
  journal={Physical review A},
  volume={36},
  number={10},
  pages={5091},
  year={1987},
  publisher={APS}
}

@article{krotov2016dense,
  title={Dense associative memory for pattern recognition},
  author={Krotov, Dmitry and Hopfield, John J},
  journal={Advances in neural information processing systems},
  volume={29},
  year={2016}
}

@article{abbott2004synaptic,
  title={Synaptic computation},
  author={Abbott, Larry F and Regehr, Wade G},
  journal={Nature},
  volume={431},
  number={7010},
  pages={796--803},
  year={2004},
  publisher={Nature Publishing Group UK London}
}

@article{kozachkov2025neuron,
  title={Neuron--astrocyte associative memory},
  author={Kozachkov, Leo and Slotine, Jean-Jacques and Krotov, Dmitry},
  journal={Proceedings of the National Academy of Sciences},
  volume={122},
  number={21},
  pages={e2417788122},
  year={2025},
  publisher={National Academy of Sciences}
}

@article{fyodorov2004complexity,
  title={Complexity of Random Energy Landscapes, Glass Transition, and Absolute Value<? format?> of the Spectral Determinant of Random Matrices},
  author={Fyodorov, Yan V},
  journal={Physical review letters},
  volume={92},
  number={24},
  pages={240601},
  year={2004},
  publisher={APS}
}

@article{franz1995recipes,
  title={Recipes for metastable states in spin glasses},
  author={Franz, Silvio and Parisi, Giorgio},
  journal={Journal de Physique I},
  volume={5},
  number={11},
  pages={1401--1415},
  year={1995},
  publisher={EDP Sciences}
}

@inproceedings{miconi2023learning,
  title={Learning to acquire novel cognitive tasks with evolution, plasticity and meta-meta-learning},
  author={Miconi, Thomas},
  booktitle={International Conference on Machine Learning},
  pages={24756--24774},
  year={2023},
  organization={PMLR}
}

@article{miconi2020backpropamine,
  title={Backpropamine: training self-modifying neural networks with differentiable neuromodulated plasticity},
  author={Miconi, Thomas and Rawal, Aditya and Clune, Jeff and Stanley, Kenneth O},
  journal={arXiv preprint arXiv:2002.10585},
  year={2020}
}

@inproceedings{miconi2018differentiable,
  title={Differentiable plasticity: training plastic neural networks with backpropagation},
  author={Miconi, Thomas and Stanley, Kenneth and Clune, Jeff},
  booktitle={International Conference on Machine Learning},
  pages={3559--3568},
  year={2018},
  organization={PMLR}
}

@article{lansner2023fast,
  title={Fast Hebbian plasticity and working memory},
  author={Lansner, Anders and Fiebig, Florian and Herman, Pawel},
  journal={Current Opinion in Neurobiology},
  volume={83},
  pages={102809},
  year={2023},
  publisher={Elsevier}
}

@article{behera2023enhanced,
  title={Enhanced associative memory, classification, and learning with active dynamics},
  author={Behera, Agnish Kumar and Rao, Madan and Sastry, Srikanth and Vaikuntanathan, Suriyanarayanan},
  journal={Physical Review X},
  volume={13},
  number={4},
  pages={041043},
  year={2023},
  publisher={APS}
}

@article{wakhloo2025associative,
  title={Associative synaptic plasticity creates dynamic persistent activity},
  author={Wakhloo, Albert J and Clark, David G and Abbott, LF},
  journal={bioRxiv},
  pages={2025--08},
  year={2025},
  publisher={Cold Spring Harbor Laboratory}
}

@article{lynch2004long,
  title={Long-term potentiation and memory},
  author={Lynch, Marina A},
  journal={Physiological reviews},
  year={2004},
  publisher={American Physiological Society}
}

@book{fischer1993spin,
  title={Spin glasses},
  author={Fischer, Konrad H and Hertz, John A},
  number={1},
  year={1993},
  publisher={Cambridge university press}
}

@book{hebb2005organization,
  title={The organization of behavior: A neuropsychological theory},
  author={Hebb, Donald Olding},
  year={2005},
  publisher={Psychology press}
}

@article{ramsauer2020hopfield,
  title={Hopfield networks is all you need},
  author={Ramsauer, Hubert and Sch{\"a}fl, Bernhard and Lehner, Johannes and Seidl, Philipp and Widrich, Michael and Adler, Thomas and Gruber, Lukas and Holzleitner, Markus and Pavlovi{\'c}, Milena and Sandve, Geir Kjetil and others},
  journal={arXiv preprint arXiv:2008.02217},
  year={2020}
}

@article{krotov2023new,
  title={A new frontier for Hopfield networks},
  author={Krotov, Dmitry},
  journal={Nature Reviews Physics},
  volume={5},
  number={7},
  pages={366--367},
  year={2023},
  publisher={Nature Publishing Group UK London}
}

@article{morris2003long,
  title={Long-term potentiation and memory},
  author={Morris, Richard GM},
  journal={Philosophical Transactions of the Royal Society of London. Series B: Biological Sciences},
  volume={358},
  number={1432},
  pages={643--647},
  year={2003},
  publisher={The Royal Society}
}

@article{dong1992dynamic,
  title={Dynamic properties of neural networks with adapting synapses},
  author={Dong, Dawei W and Hopfield, John J},
  journal={Network: Computation in Neural Systems},
  volume={3},
  number={3},
  pages={267},
  year={1992},
  publisher={IOP Publishing}
}

@article{vyas2020computation,
  title={Computation through neural population dynamics},
  author={Vyas, Saurabh and Golub, Matthew D and Sussillo, David and Shenoy, Krishna V},
  journal={Annual review of neuroscience},
  volume={43},
  number={1},
  pages={249--275},
  year={2020},
  publisher={Annual Reviews}
}

@article{little1974existence,
  title={The existence of persistent states in the brain},
  author={Little, William A},
  journal={Mathematical biosciences},
  volume={19},
  number={1-2},
  pages={101--120},
  year={1974},
  publisher={Elsevier}
}

@article{gardner1988optimal,
  title={Optimal storage properties of neural network models},
  author={Gardner, Elizabeth and Derrida, Bernard},
  journal={Journal of Physics A: Mathematical and general},
  volume={21},
  number={1},
  pages={271},
  year={1988},
  publisher={IOP Publishing}
}

@article{clark2024dynamic,
  title = {Theory of Coupled Neuronal-Synaptic Dynamics},
  author = {Clark, David G. and Abbott, L. F.},
  journal = {Phys. Rev. X},
  volume = {14},
  issue = {2},
  pages = {021001},
  numpages = {20},
  year = {2024},
  month = {Apr},
  publisher = {American Physical Society},
  doi = {10.1103/PhysRevX.14.021001},
  url = {https://link.aps.org/doi/10.1103/PhysRevX.14.021001}
}

@article{hopfield1982,
  title={Neural networks and physical systems with emergent collective computational abilities},
  author={Hopfield, John J},
  journal={Proceedings of the national academy of sciences},
  volume={79},
  number={8},
  pages={2554--2558},
  year={1982},
  publisher={National Acad Sciences}
}

@article{hopfield1984,
  title={Neurons with graded response have collective computational properties like those of two-state neurons},
  author={Hopfield, John J},
  journal={Proceedings of the National Academy of Sciences},
  volume={81},
  number={10},
  pages={3088--3092},
  year={1984},
  publisher={National Acad Sciences}
}

@article{amit1985,
  title={Storing infinite numbers of patterns in a spin-glass model of neural networks},
  author={Amit, Daniel J and Gutfreund, Hanoch and Sompolinsky, Haim},
  journal={Physical review letters},
  volume={55},
  number={14},
  pages={1530},
  year={1985},
  publisher={APS}
}

@article{amit1987,
  title={Statistical mechanics of neural networks near saturation},
  author={Amit, Daniel J and Gutfreund, Hanoch and Sompolinsky, Haim},
  journal={Annals of physics},
  volume={173},
  number={1},
  pages={30--67},
  year={1987},
  publisher={Elsevier}
}

@book{hebb1949,
  title={The organization of behavior: A neuropsychological theory},
  author={Hebb, Donald Olding},
  year={1949},
  publisher={Wiley}
}

@article{tsodyks1997,
  title={The neural code between neocortical pyramidal neurons},
  author={Tsodyks, Misha V and Markram, Henry},
  journal={Proceedings of the National Academy of Sciences},
  volume={94},
  number={2},
  pages={719--723},
  year={1997},
  publisher={National Acad Sciences}
}

@article{mongillo2008,
  title={Synaptic theory of working memory},
  author={Mongillo, Gianluigi and Barak, Omri and Tsodyks, Misha},
  journal={Science},
  volume={319},
  number={5869},
  pages={1543--1546},
  year={2008},
  publisher={American Association for the Advancement of Science}
}

@article{sompolinsky1988,
  title={Chaos in random neural networks},
  author={Sompolinsky, Haim and Crisanti, Andrea and Sommers, Hans-J{\"u}rgen},
  journal={Physical review letters},
  volume={61},
  number={3},
  pages={259},
  year={1988},
  publisher={APS}
}

@book{mezard1987,
  title={Spin glass theory and beyond},
  author={Mezard, Marc and Parisi, Giorgio and Virasoro, Miguel Angel},
  year={1987},
  publisher={World Scientific Publishing Company}
}

@misc{clark2025transient,
  title={Transient dynamics of associative memory models},
  author={Clark, David G},
  year={2025},
  eprint={2506.05303v2},
  archivePrefix={arXiv},
  primaryClass={cond-mat.dis-nn}
}

@article{marsh2021enhancing,
  title={Enhancing associative memory recall and storage capacity using confocal cavity QED},
  author={Marsh, Brendan P and Guo, Yudan and Kroeze, Ronen M and Gopalakrishnan, Sarang and Ganguli, Surya and Keeling, Jonathan and Lev, Benjamin L},
  journal={Physical Review X},
  volume={11},
  number={2},
  pages={021048},
  year={2021},
  publisher={APS}
}

@article{marsh2025high,
  title={High-capacity associative memory in a quantum-optical spin glass},
  author={Marsh, Brendan P and Schuller, David Atri and Ji, Yunpeng and Hunt, Henry S and Ganguli, Surya and Gopalakrishnan, Sarang and Keeling, Jonathan and Lev, Benjamin L},
  journal={arXiv preprint arXiv:2509.12202},
  year={2025}
}
\end{document}